\RequirePackage{fix-cm}
\documentclass[smallextended]{svjour3}       
\usepackage{graphicx}
\usepackage{mymacros}
\usepackage[textsize=scriptsize,prependcaption]{todonotes}

\begin{document}

\title{
Tasking framework for Adaptive Speculative Parallel Mesh Generation
}

\author{Christos Tsolakis \and Polykarpos Thomadakis \and Nikos Chrisochoides}

\institute{
Christos Tsolakis, Polykarpos Thomadakis, Nikos Chrisochoides \\
Center for Real-Time Computing, Old Dominion University, Virginia, USA\\
\email{ \{ctsolakis,pthomadakis,nikos\}@cs.odu.edu}
}

\maketitle

\begin{abstract}
Handling the ever-increasing complexity of mesh generation codes along
with the intricacies of newer hardware often results in codes that are both 
difficult to comprehend and maintain.
Different facets of codes such as thread management  and load balancing
are often intertwined, resulting in efficient but highly complex software.
In this work, we present a framework which aids in establishing
a core principle, deemed \emph{separation of concerns},
where \emph{functionality} is separated from 
\emph{performance} aspects of various mesh operations.
In particular, thread management and scheduling decisions are elevated into 
a generic and reusable tasking framework.

The results indicate that our approach
can successfully abstract the load balancing 
aspects of two case studies,  while
providing access to
a plethora of different execution back-ends. 
One would expect, this new flexibility to lead to some additional cost.
However, 
for the configurations studied in this work,
we observed up to $13\%$ speedup
for some meshing operations and up to $5.8\%$ speedup 
over the entire application runtime  compared to hand-optimized code.
Moreover, we show that by using different task creation strategies, the
overhead compared to straight-forward task execution models
can be improved dramatically by as much as
$1200\%$ without compromises in portability and functionality.

\end{abstract}


\section{Introduction}

With the advent of multicore machines and new constantly evolving architectures 
in the form of accelerators, the need for abstracting parallelism in scientific 
applications becomes paramount. 
Although platform-specific and/or application-specific  optimizations will
always perform better than generic solutions, abstract interfaces can last 
longer and allow for better interoperability between applications.
Choosing the right abstractions allows applications to build upon a generic
framework while enabling low-level software substrates to offer implementations that 
take advantage of the underlying hardware. Moreover,
abstractions provide space 
for future explorations and allow to
future-proof~\cite{furrer_future-proof_2019} applications; as newer hardware
(e.g., in the form of accelerators) 
becomes available, the application developer may need to perform minimal to 
no changes while the underlying runtime system can add new features opaquely.

\begin{figure}[!htpb]
	\centering
	\begin{subfigure}[!htpb]{0.465\linewidth}
	\includegraphics[width=\linewidth]{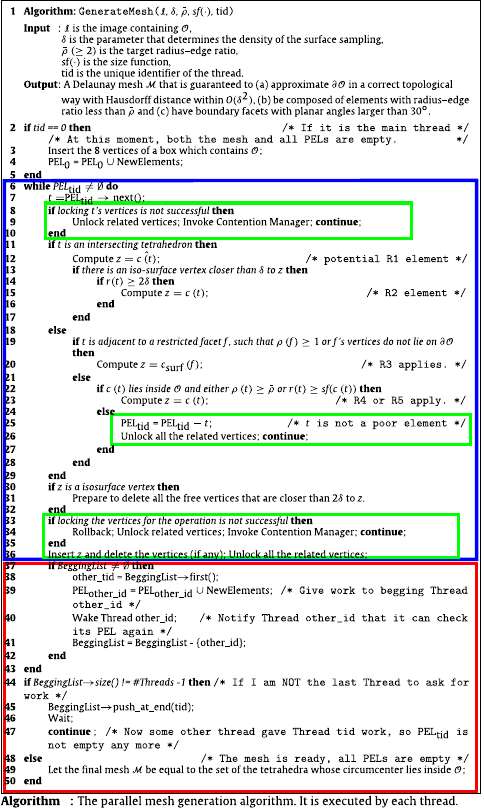}
	\caption{PODM pseudocode as presented in \cite{foteinos_high_2014}.}
	\label{fig:podm_code}
	\end{subfigure}
	\begin{subfigure}[!htpb]{0.49\linewidth}
	\includegraphics[width=\linewidth]{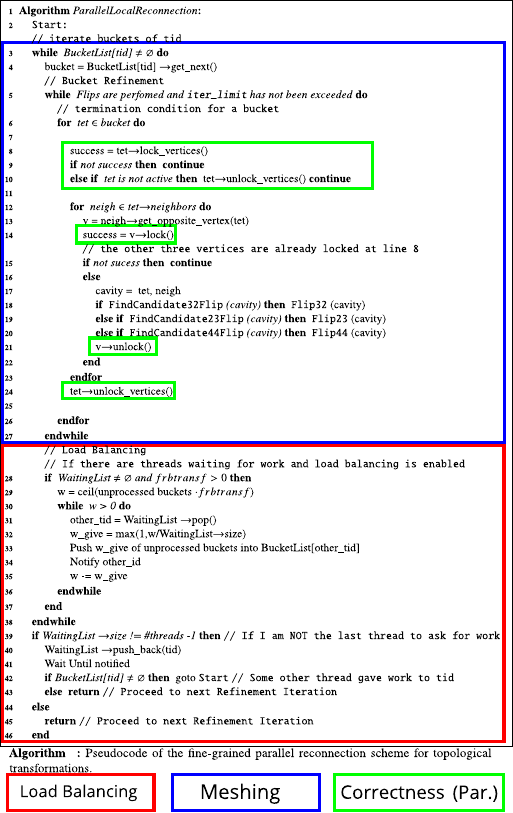}
	\caption{CDT3D pseudocode as presented in 
	\cite{drakopoulos_fine_grained_2019}.}
	\label{fig:cdt3d_code}
	\end{subfigure}
	\caption{Pseudocodes of the speculative approach applied to a 
	Delaunay-based algorithm  (left) and a local reconnection operation  
	(right) of an Advancing-Front method.
	Colored regions indicate the primal function of the enclosed steps.}
	\label{fig:codes}
\end{figure}

The task management facilities presented in this work, are
part of a longer-term project that aims to enable \emph{separation of
concerns}~\cite{dijkstra_role_1982}  with regard to \emph{functionality} and \emph{performance}, 
specifically for mesh operations. Figure \ref{fig:codes} depicts the 
pseudocode of two fine-grained speculative meshing operations. Note 
that the application developer must manage and account for 
the meshing  kernel, parallel correctness, and 
load balancing, all within a single  algorithm.
Developing and maintaining such an application becomes challenging 
since the developer has to keep all three parts in mind 
while modifying the code. 
Moreover, re-using hardware-specific optimizations among different applications
can only be achieved  by abstracting them outside of specific applications.
Examples include data affinity-aware work schedulers, cache-optimized data
structures, thread contention management policies, etc.

The separation of \emph{functionality} from  \emph{performance}
will contribute  significantly  towards the 
implementation 
of the \emph{Telescopic Approach} \cite{chrisochoides_telescopic_2016} which 
lays down a design for achieving scalability for mesh generation on exascale
machines. 
The \emph{Telescopic Approach} spans across the multiple 
memory  hierarchies of an exascale machine
(shared, distributed-shared (DSM), distributed, out-of-core) 
and maps different algorithmic layers to the appropriate level of memory
based on the intensity of communication between different meshing kernels.
The lowest, closest to the hardware layer, 
that can sustain high volume of communication at low cost,
is the Parallel Optimistic layer which 
is designed to explore concurrency at the CPU level within the limits of shared 
memory, using 
speculative/optimistic execution. Both meshing kernels of Figure 
\ref{fig:codes} are designed to be used in the Parallel Optimistic layer
of the  \emph{Telescopic Approach}.
Abstracting the \emph{performance} aspects from the \emph{functionality} 
for these kernels will allow interoperability with lower level runtime 
systems 
like PREMA  \cite{Thomadakis20M}
and will speed up the development process
by increasing the code-reuse among the applications that utilize the
\emph{Telescopic Approach}.

As case studies, we use the parallel meshing operations present in CDT3D 
\cite{drakopoulos_fine_grained_2019,drakopoulos_finite_2017}
that are common in most metric-based mesh adaptation codes 
\cite{tsolakis_parallel_2019} and the Delaunay-based kernel of PODM \cite{foteinos_high_2014}.
For these two applications, we explore 
ways a tasking environment 
can be used to express speculative mesh operations
and describe approaches that enable to abstract the load balancing aspects of both
case studies with no impact to functionality.
The results in Sections \ref{sec:case-study-CDT3D} and 
\ref{sec:case-study-PODM} indicate not only low overhead, but even 
speedup with respect to the baseline hand-optimized applications
for some mesh operations.
In summary, the main contributions of this study are:
\begin{itemize}
    \item Present a high-level front-end that abstracts and
    unifies task management for adaptive and irregular applications.
    \item Design and implement the front-end
	for three major back-ends: Intel\R's TBB, OpenMP, and Argobots.
    \item Illustrate how this front-end can be applied to two
    different speculative parallel unstructured mesh
    generation codes. 
    \item Provide an in-depth analysis of the effect of task granularity on
    each back-end along with the advantages and disadvantages of different
     task creation strategies.
\end{itemize}

\section{Related Work}
\label{sec:speculative-approach}
There is a number of tasking systems 
that have emerged in the recent years targeting different layers of the application,
from low-level assembly layer \cite{rainey_task_parallel_assembly_2021},
and parallelizing compilers \cite{ying_T4_compiling_2020} to high level
tasking frameworks \cite{chi_extending_high-level_synthesis_2021}.
A complete review of the current state-of-the-art tasking environments is 
outside the scope of this section. A comprehensive taxonomy based on
architectural characteristics and user APIs appears in~\cite{thoman_taxonomy_2018}.
In the rest of this section, we focus on methods that exploit concurrency 
through speculative execution.
\emph{Speculative} execution (also known as \emph{optimistic}) is a technique
used in a number of applications, ranging from processors 
\cite{tomasulo_efficient_1967} to databases \cite{kung_optimistic_1981}. 
It allows for the exploitation of more concurrency out of a problem by 
executing steps of a procedure ahead of time, prior to resolving data
dependencies between the steps themselves.
In the case where steps are not 
needed,  the precomputed results may either be disregarded or additional steps may be required to \emph{roll back} to the
previous state of the process. 
The correctness of this scheme, in the context of parallel
algorithms, has been proven in~\cite{jefferson_virtual_1985} with the 
introduction of the notion of \emph{Virtual Time} and validated in the 
context of Parallel Discrete Event Simulations within the 
\emph{Time Warp} system.

There are several efforts in the literature that facilitate speculative execution 
utilizing higher level constructs. Among the many we list a few pertinent to this study such as the use of 
transactional memory at the software level~\cite{raman_speculative_2010}, 
compiler-assisted methods~\cite{rauchwerger_lrpd_1995,caamano_apollo_2017,openmp_speculation}
and libraries such as Galois~\cite{kulkarni_optimistic_2007}, ParlayLib~\cite{blelloch_parlaylib_2020}
and SPETABARU~\cite{bramas_increasing_2019}.

The Galois system~\cite{kulkarni_optimistic_2007} 
provides abstract set iterators, giving to the application developer
the ability to extract parallelism out of the work-lists of a sequential application. Custom data 
structures and a runtime scheduler are responsible for detecting
and recovering unsafe accesses to shared memory. 
The elegance of this approach is appealing but, for our use-case, it
would  require extensive modifications of the work lists maintained by each
application. Also, to the best of our knowledge, its effectiveness for mesh
generation has been demonstrated only on simple
sequential mesh triangulation
codes. 
In contrast, in this work, both use-cases build on top of an
already-parallelized application that have demonstrated comparable 
performance to state-of-the-art methods~\cite{foteinos_high_2014,tsolakis_parallel_2021}.

In~\cite{blelloch_internally_2012,blelloch_parlaylib_2020} the authors revisit 
the idea of expressing
speculative execution as a combination of nested parallelism and commutative 
operations suggested in~\cite{steele_making_1989} and propose 
the use of \emph{deterministic reservations} for dealing with a class of greedy
algorithms. The main idea is to split the 
operation into two phases. 
One that attempts to reserve the data dependencies for
a number of tasks speculatively, and then a commit phase 
that executes the tasks that successfully reserved 
all their dependencies.
The two-phase approach is similar to the 
\emph{inspector-executor} model~\cite{saltz_runtime_1991}.
However, an
\emph{inspector-executor} model is not suitable for
our case due to the data-intensive nature of the targeted use-cases.
In contrast, our approach
re-uses the speculative
approach step already present in both use-cases and merges the two steps
in one. This approach acts directly upon touched data which improves
cache utilization and allows tolerating more than $80\%$ of system latencies~\cite{nave_guaranteed_2002}.
Moreover, it avoids the synchronization required by a two-phase approach.

The SPETABARU tasking runtime system introduced in~\cite{bramas_increasing_2019}
can exploit concurrency of task graphs 
through speculation. The task graph is built based on the user-defined data 
dependencies between the tasks. The system manages the execution of tasks 
as well as disregarding data from failed speculative attempts upon runtime.
The library was originally created for Parallel Monte Carlo Simulations, 
and it is primarily designed for parallel applications that utilize graphs
of tasks. 
This tasking system generates the graph utilizing a single thread in a pre-processing step
that generates all the tasks and evaluates the data dependencies among them.
This approach is inadequate for our target data-intensive applications for two reasons.
First, dependency discovery and resolution is the most expensive step, thus rendering
the pre-processing step to a major bottleneck. Moreover, the continuous generation of 
new elements (and therefore tasks) would require additional synchronization points
which would degrade performance significantly.

In~\cite{openmp_speculation} the authors incorporate Thread-Level Speculation
in OpenMP.  
Open\-MP is extended with the \texttt{speculative} directive which 
annotates a variable as the target of speculative execution. A thread-local
version of such variables is created for each thread. The respective 
runtime monitors such variables and guarantees that all read accesses will
return the most up-to-date value. When a thread consumes an outdated version of
a \texttt{speculative} variable, it is stopped and restarted in order to 
consume the correct value of the variable. 
For both of our applications the speculative execution is already
part of the application code and modifying it is outside the 
scope of this work. Moreover, the abstract front-end
of our approach gives access to additional back-ends beyond OpenMP.

Although many of the above approaches are close to our goals,
most of them will require nontrivial changes to the code
required for the
\emph{Parallel Correctness} steps (see Figure~\ref{fig:codes})
of the algorithm which for this study we chose to keep as
part of the meshing task. 
Moreover, in contrast to all the
presented approaches, the starting point in this work is applications
that are already parallel instead of sequential. 
This comes with the benefit of having thread-safe
mechanisms in place for memory allocation
and speculative locking, but also
with higher complexity due to their legacy nature.
Also, the proposed approach  can utilize a number of different back-ends,
including Argobots~\cite{seo_argobots_2018}.
Argobots provides lightweight User Level Threads (ULTs),
capable of context switching with low overhead. Among others, ULTs
are employed to tolerate latencies in cases such as failing to acquire
a lock or calling synchronous MPI operations where regular tasks
would block, along with the underlying hardware thread. Instead, 
a ULT will implicitly release the hardware thread it runs on, 
allowing other ULTs to use it.
This interaction with MPI allows to build scalable runtime systems such as the PREMA
runtime system~\cite{Thomadakis20M}, which in turn is a building block of the
\emph{Telescopic Approach}~\cite{chrisochoides_telescopic_2016}.

In the context of mesh generation, speculative execution  was introduced 
in~\cite{nave_guaranteed_2002} 
where the authors execute the same meshing kernel 
across multiple processes without  restricting them on their local data. 
Instead, the meshing kernel  is launched \emph{optimistically}, and 
the data dependencies are discovered and captured on  the fly. 
If some dependency cannot be satisfied, the operation releases any 
captured  dependencies, aborts its execution (rollback), and reenters 
the scheduler's pool. 
The speculative approach has already been utilized in Delaunay 
triangulation methods~\cite{blandford_engineering_2006,batista_parallel_2010,foteinos_dynamic_2011}, 
Delaunay Refinement~\cite{foteinos_high_2014} and Advancing-Front ~\cite{drakopoulos_fine_grained_2019} methods. 
However, in each case, the approach was application- 
and method-specific.
In contrast, in this work we provide an abstract interface
allowing the framework to be applied both across different 
mesh operations and between different mesh applications.
Moreover, the approach in this work is method-agnostic, thus
rendering the framework useful for other meshing methods as well as for
adaptive and irregular applications in general.

\section{Method }

The proposed approach builds upon the observation
that the speculative meshing operations of the two case 
studies can be decomposed  into three components: 
\emph{Meshing}, \emph{Parallel Correctness} and \emph{Load Balancing}
(see Figure~\ref{fig:codes}).
The Load  Balancing part is handled by the generic 
tasking framework presented in the following sections.
\emph{Parallel Correctness} is expressed through the use of atomic 
locks upon the cavity (data dependencies) of each operation.
For this study, the parallel correctness steps will remain  as
part of the meshing task.
The implementation of generic tasking framework is composed of an
application-facing front-end which is agnostic to target hardware, and a
back-end, which provides custom implementations for individual substrates.

\subsection{Front-end: A Generic Tasking Framework}
\label{sec:general-tasking-framework}

\begin{figure}[!htbp]
\centering
\includegraphics[width=0.3\linewidth]{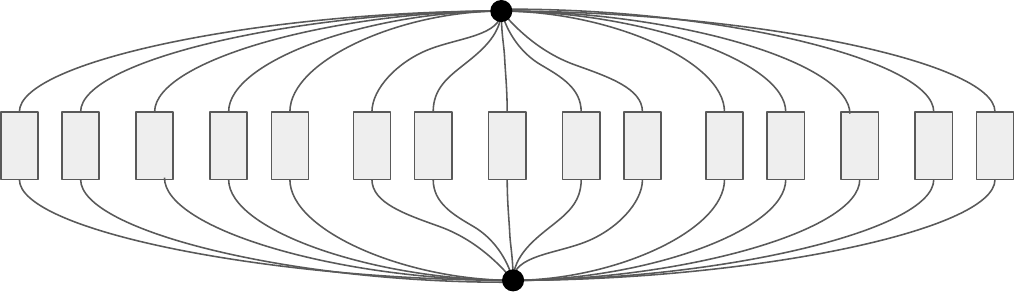}
\hfill
\includegraphics[width=0.3\linewidth]{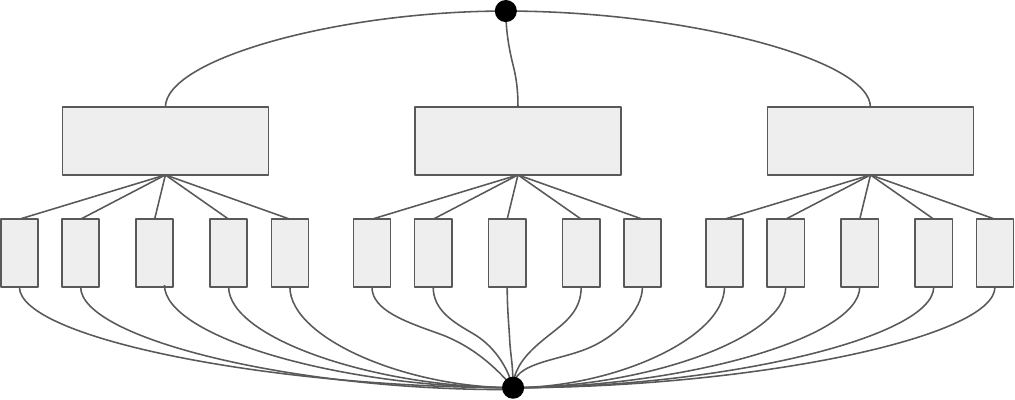}
\hfill
\includegraphics[width=0.2\linewidth]{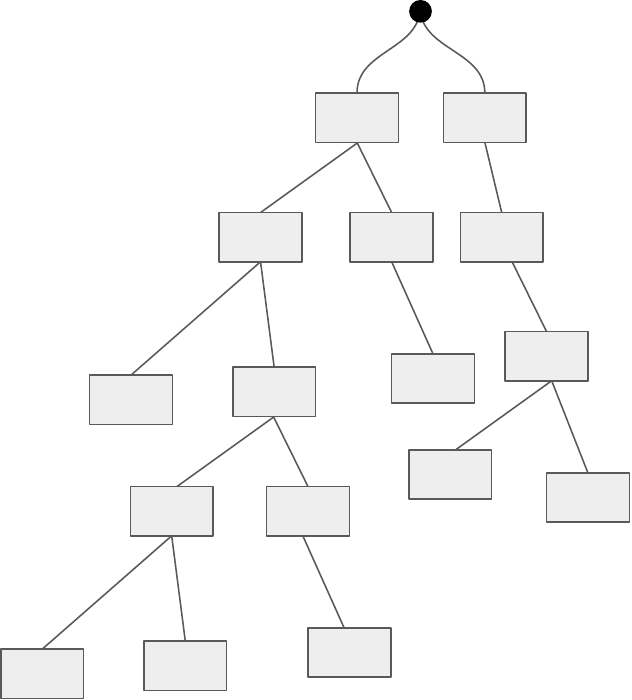}
\caption{Different tasking paradigms employed in this work. Left: flat 
	model, Middle: two-level task creation, Right: hierarchical task creation.}
\label{fig:task_specturm}
\end{figure}

The general approach used in this work is to decompose any given operation into 
non-interruptible  tasks that execute to completion. 
The framework supports both blocking calls for creating many 
tasks (see Listing~\ref{code:task_for}) and a fine-grained API for single task
creation (see Listing~\ref{code:create_task}). This allows for the 
utilization of a range of tasking paradigms, from simple fork-join 
models to hierarchical or recursive task creation
(see Figure~\ref{fig:task_specturm}).

\begin{lstlisting}[
style=my_cpp_style,
caption={Interface for launching several tasks at once.},
label=code:task_for	
]
/**
* @brief launch tasks and wait until they all finish
* @param user_task_args vector of task arguments
* @param user_func the user function to be executed, 
* type should be void func(UserTaskArgs&), 
* @param grainsize number of elements of user_task_args to group 
* into a single task
*/
	
template<typename UserTaskArgs, typename FunctionType>
void task_for(std::vector<UserTaskArgs>& user_task_args, 
              FunctionType user_func, 
              int grainsize = 10)
\end{lstlisting}

\begin{lstlisting}[
	style=my_cpp_style,
	caption={Interface for creating a single task.},
	label=code:create_task
	]
/**
* @brief add a task to the internal queue
* @param user_task_args arguments of the task
* @param user_func the user function to be executed, 
* type should be void func(UserTaskArgs&) 
*/

template<typename UserTaskArgs,typename FunctionType>
void create_and_schedule(UserTaskArgs& user_task_args, 
                         FunctionType user_func)

/**
* @brief wait until all generated tasks have completed.
*/
void wait_for_all()
\end{lstlisting}

\texttt{task\_for} in Listing~\ref{code:task_for} is similar to the 
parallel-for paradigm.
It accepts a function \texttt{user\_func} 
that implements the task operation, as well as a vector \texttt{user\_task\_args} of the different 
arguments for each task. Optionally, it can accept a grainsize which
controls the number of terminal tasks  that will 
be generated. The number of terminal tasks is 
\texttt{user\_task\_args.size()/grainsize}.
Similar to \texttt{std::for\_each}, \texttt{task\_for} will apply
\texttt{user\_func} to each element of the  
\texttt{user\_task\_args} vector.
However, in contrast to \texttt{std::for\_each} not all invocations of  
\texttt{user\_func} will be completed successfully. Some will abort due to rollbacks. In this study, re-applying the operation on aborted tasks is handled 
by the application logic, since it was already present before the introduction
of this framework. 

\texttt{create\_and\_schedule} in Listing~\ref{code:create_task} is a simple 
wrapper around the corresponding 
back-end that generates a task and places it in the internal queue of the 
framework. This call is not blocking, and the execution of the task may start 
immediately on a different thread. Finally, \texttt{wait\_for\_all} suspends 
the 
calling thread until the internal task queues are empty.

The tasking framework also provides the user with a unique \texttt{thread\_id} 
$\in [0,nthreads)$. This id is not pinned to any hardware thread, but it is 
guaranteed to stay fixed and unique for the duration of the task execution.
The \texttt{thread\_id} is required by both applications of this study for 
two main operations. First, data dependency acquisition is implemented 
utilizing atomic locks that hold the id of the owning thread~\cite{foteinos_high_2014,drakopoulos_fine_grained_2019}. Also, 
both pieces of software utilize the thread-aware memory management method 
described 
in~\cite{antonopoulos_multigrain_2005} that uses a \texttt{thread\_id} in 
order to access the appropriate thread-local memory pools that allow for
the allocation and deallocation of elements in a thread-safe manner.

\subsection{Task Generation Strategies}

One of the considerations of explicitly creating tasks is 
the overhead of task creation. In the current implementation
of \texttt{task\_for}, there is support for all three task creation strategies
of Figure~\ref{fig:task_specturm}.
The \texttt{flat} model implements a basic fork-join paradigm~\cite{conway_multiprocessor_1963}. It creates all tasks sequentially and waits 
for them to complete. As a first attempt to reduce the overhead, 
a \texttt{2level} task creation strategy was introduced. For this strategy,  
the application thread will spawn sequentially $2 \cdot nthreads$ tasks that 
partition the range of the
\texttt{user\_task\_args} vector in equal parts. Each level-2 task will then 
iterate the assigned range of the task vector and spawn a task for each 
task-argument. 
Finally,  the \texttt{hierarchical} model employs a divide-and-conquer scheme;
it creates tasks recursively by creating two child 
tasks that bisect the task range up to the point where the assigned range is 
smaller or equal to the  target grainsize. 
When the framework is used sequentially, no tasks are created independently
of the chosen strategy. Instead, the application thread will apply 
\texttt{user\_func} sequentially on each item of the \texttt{user\_task\_args}
vector.

\subsection{Implementation}

The above framework is implemented with three different 
back-ends: the Argobots runtime system~\cite{seo_argobots_2018}, 
Intel’s TBB framework~\cite{intel_tbb_2008} and OpenMP~\cite{openmp_1998}. 
For each of the three implementations we have incorporated the three task creation strategies
of Figure~\ref{fig:task_specturm} as well as high level constructs specific to each
back-end such as \tbbfor,
\ompfor and \omptaskloop for a total of $12$ different execution back-ends.
We will use the notation \texttt{backend}-\texttt{strategy} to refer 
to tasking strategy \texttt{strategy} implemented on top of the back-end \texttt{backend}.
Although interoperability in terms of mixing different backends is possible, for this 
work we restrict our attention to using one back-end at a time.

\subsubsection{Argobots back-end implementation details} \label{sec:argobots-details}
Argobots is a low-level tasking framework developed to support higher level 
runtime systems, so it does not provide optimized schedulers for fork-join parallelism
out of the box. To implement an optimized tasking framework for our needs, 
we developed custom scheduling mechanisms  using the  interfaces
provided. For this work we opted for a  work-stealing \cite{blumofe_scheduling_1999}
scheduling mechanism
even though any other mechanism could be incorporated in the future.
In this scheduling mechanism, each thread (execution stream in Argobots' terminology) in the 
parallel environment is associated with a circular double-ended queue (deque) 
which is thread-safe and lock-free~\cite{lev_deque}. 
Every new task is pushed to the top of the deque of the thread 
that 
created it. When a thread finishes with the
execution of a task, it first checks its own deque; if there are tasks available, it pops the one residing at the top of the deque and executes it. If its deque
is empty, it will randomly pick one of the remaining threads and try to steal the 
task at the bottom of its deque. By picking the task at the top of the owned deque
first, tasks that are hot in the cache are given priority. On the other hand,
stealing the task at the bottom of other threads' deques: increases the 
chance of picking tasks that will create more child tasks,
allows more work to become available for the stealing thread and results in a 
decreased number of steal attempts. We provide two tasking flavors for this implementation - User
Level Threads (ULTs) that can yield explicitly and Tasklets that run to completion
and can only block waiting for another tasklet created using this framework.
In this work, both case studies use tasklets as we only need to wait for other 
tasks
to complete and no other blocking operation is performed. Each task is created
using a  \texttt{abt::task\_create} function call that asynchronously schedules 
a new task and immediately returns a task handle. The task handle can then be 
used to check or wait for the completion of the respective task's execution. 
Internally, the call to wait for a task completion will result in calling the 
scheduler and popping/stealing some other task. 

\subsubsection{TBB back-end implementation details} \label{sec:tbb-details}
Intel\R Thread Building Blocks (TBB)\footnote{Recently, Intel\R Threading Building Blocks was renamed to Intel\R 
oneAPI Threading Building Blocks (oneTBB) to highlight that the tool is part of 
the oneAPI ecosystem.}
is a library that enables parallel programming across different applications and 
architectures. It provides high level constructs such as \tbbfor 
in addition to giving access to the lower level tasking queues. TBB uses tasks to express parallelism,
thus making it a suitable candidate for this study. Tasks are expected to be 
non-preemptive, which is the case for both applications of this study and for 
speculative 
operations in general. The scheduler switches the running thread only when a 
task is waiting for its spawned children.
For the \texttt{hierarchical} and the  \texttt{2level} task creation strategy, 
each level is enclosed in a \texttt{tbb::task\_group} that allows to wait until all tasks 
of the group  are completed. When using the lower level 
\texttt{create\_and\_schedule}, all generated tasks are added to the same 
global \texttt{tbb::task\_group} thus allowing termination to be detected in a
convenient manner while still enabling work stealing among all threads. 
Additionally, we 
implemented a wrapper that passes the arguments of \texttt{task\_for}
directly to the higher level  \texttt{tbb::parallel\_for} function,
in order to compare it with our framework.

\subsubsection{OpenMP back-end implementation details} \label{sec:openmp-details}
OpenMP is an API that enables parallel shared-memory programming with the use of 
\texttt{\#pragma}s making it easily accessible directly through the compiler. 
It is included in this study since it is often the first step towards introducing
parallelism for many scientific applications.
Tasks are created using \texttt{\#pragma omp task} and they are
declared as \texttt{untied} which gives them the opportunity to be scheduled
 on any available thread. For the \texttt{hierarchical}
strategy,  it was advantageous to prepend \texttt{\#pragma omp taskyield}
right before the recursive step. This created an extra scheduling 
opportunity for the back-end. Without it, it was noticed that
a single thread would tend to run all the tasks it created, affecting performance
and greatly increasing the recursion tree size.
For comparison, we also implemented two more wrappers using higher level 
constructs. The first passes the arguments of 
\texttt{task\_for} directly to  
\texttt{\#pragma omp for} while the second passes to  
\texttt{\#pragma omp taskloop}. For the \texttt{\#pragma omp for}, we chose the 
\texttt{dynamic} scheduler because it performs on average better across
the different mesh operations covered in this work.

\section{Case Studies}

As case studies we use the parallel meshing operations present in 
CDT3D ~\cite{drakopoulos_fine_grained_2019,drakopoulos_finite_2017}
and the Delaunay-based kernel of PODM~\cite{foteinos_high_2014}.
These two applications share a lot of common ideas when it comes to parallel
execution, but they also have some differences. 
As mentioned in Figure~\ref{fig:codes}, both applications utilize speculative
execution for their meshing operations which they implement similarly;
the meshing kernel (blue sections) uses atomic locks speculatively to guarantee 
correctness (green sections).
This feature fits well with our approach since we assume that each task should be 
non-interruptible and should execute to completion.
Also, they both integrate load balancing and thread
management with the mesh application (red sections),
that our framework can abstract away.

PODM is built around a single mesh operation for modifying the mesh, thus reducing the
amount of code changes required.
On the other hand,
when it comes to parallel execution, it has a number of optimizations that 
complicate the use of the tasking framework. In particular, 
it uses a Hierarchical Load Balancing that ties worklists to specific threads
in order to improve data affinity and takes into account the cost of memory access 
when moving load between threads. These optimizations result in a 
tight coupling between the mesh operations and the load balancing parts of the code.

In contrast, in CDT3D  the coupling between the threads and their data is lower.
At a high level it follows a fork-join pattern where sequential steps prepare 
global data structures
for parallel execution. This structure matches well with 
the \texttt{task\_for}  API and reduces the
places where the code needs to be modified.
Moreover, it utilizes a set of different mesh operations that use
both hand-optimized and generic work sharing methods,
which results  in varying impact on performance when 
transitioning to the tasking framework.

\subsection{ Case Study I : Parallel Mesh Adaptation Software
(CDT3D)}\label{sec:case-study-CDT3D}
\begin{figure}[tbph!]
	\centering
	\includegraphics[width=0.5\linewidth]{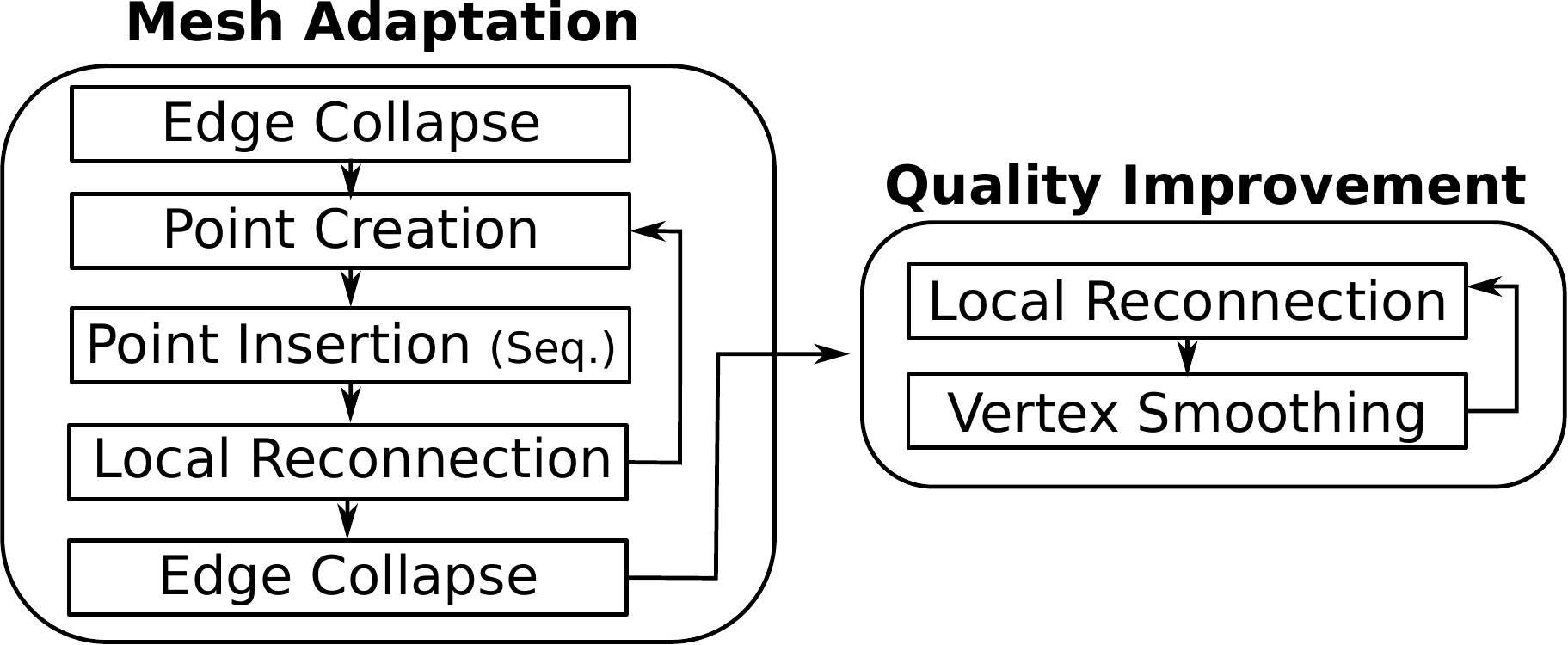}
	\caption{ Mesh operations in CDT3D.}
	\label{fig:mesh_operations}
\end{figure}

CDT3D is composed of many modules depicted in 
Figure~\ref{fig:mesh_operations}. With proper re-arrangement of the modules
one could
implement different meshing applications as described 
in~\cite{christos_tsolakis_exascale-era_2020}. The configuration
chosen for this study is optimized for metric-based adaptation and
has been already compared against state-of-the-art mesh adaptation
codes in~\cite{tsolakis_parallel_2021}.

For this case-study, the focus is on: Point Creation, Local Reconnection, Edge 
Collapse, and Vertex Smoothing. 
The common first step for porting the operations is to express them in a way that is 
compatible with 
the API of the front-end presented in Figure~\ref{code:task_for}.
The most natural choice is as an operation applied to an element. However, the
baseline implementation of CDT3D already uses ``buckets'' (i.e, lists of
elements) for some of its operations (see, for example, Figure
\ref{fig:cdt3d_code} and references~\cite{drakopoulos_fine_grained_2019,drakopoulos_finite_2017}).
In the context of the presented mesh operations, ``buckets'' are used as simple
strip partitioning method similar to the \texttt{chunk-size} parameter of the
\ompfor scheduler.
In an effort to maximize code re-use of the application, we opted for the 
conventions of Table~\ref{tab:operation-characteristics}.

\begin{table}[!htpb]
			\caption{Characteristics of the baseline implementation of the parallel 
		mesh operations ported to the tasking framework.}
{
\footnotesize
\centering
\begin{tabular}{lp{1.7cm}p{3cm}l}
\textbf{Operation}  & \textbf{Work-unit}        & \textbf{Operator} & 
\textbf{Baseline implementation} \\ \hline
Local Reconnection  & ``Bucket'' & Apply local reconnection between 
an element and its face neighbors for each element of a bucket & custom 
scheduling (\cite{drakopoulos_fine_grained_2019})
\\
Point Creation      & ``Bucket'' & Generate candidate 
points for each element of the bucket              & custom scheduling 
(\cite{drakopoulos_fine_grained_2019}) \\
Edge Collapse       & Vertex                       & Collapse small edges 
attached to a mesh vertex              & \texttt{omp for 
schedule(guided)}              \\
Vertex Smoothing    & Vertex                       & Improve the quality of the 
elements attached to a mesh vertex by smoothing & \texttt{omp for 
schedule(static)} 
	\end{tabular}
}
	\label{tab:operation-characteristics}
\end{table}

One important feature of the \emph{Operator} in each case, is that it is built 
using the speculative/optimistic approach.  In practice, it means that no 
data or domain decomposition is applied to the mesh, but the operator will 
attempt to acquire its dependencies through some exclusive locking mechanism 
upon execution. Failure to do so will result in unlocking any acquired 
resources and exiting.

\subsubsection{Performance Evaluation}\label{sec:cdt3d-evaluation}
For this evaluation, the code was recompiled picking the appropriate back-end
implementation 
each time. The experiments were performed on the \texttt{wahab} cluster of Old Dominion 
University using  dual socket nodes equipped with
two Intel\R Xeon\R 
Gold 6148 CPU @ 2.40GHz (20 slots) and  368 GB of memory.
The compiler is \texttt{gcc 7.5.0} and the compiler flags \texttt{-O3 -DNDEBUG -march= native}. \texttt{gcc 7.5.0} comes with support of OpenMP version \texttt{4.5}. For TBB, version \texttt{2021.1.1} was used. Each configuration was executed 10 times. 
All times are normalized based on the performance of the baseline application,
unless it is stated otherwise.
The graphs below use the geometric mean~\cite{fleming_how_1986} to summarize 
the results for each configuration. 
The evaluation in the following paragraphs proceeds as follows:
First, we compare higher-level constructs (\ompfor, \omptaskloop and \tbbfor)
to the \texttt{-flat} strategy implemented using the three back-ends.
Next, we compare the \texttt{-flat}, \texttt{-2level}, and
\texttt{-hierarchical} strategies as implemented in our framework.
We then analyze and optimize the grainsize for each back-end and strategy
in order to derive the optimal grainsize for each operation. Finally,
we compare our framework using the optimal grainsizes with the baseline
application.

\begin{figure}[!htpb]
	\centering
	
	\includegraphics[width=0.49\linewidth]{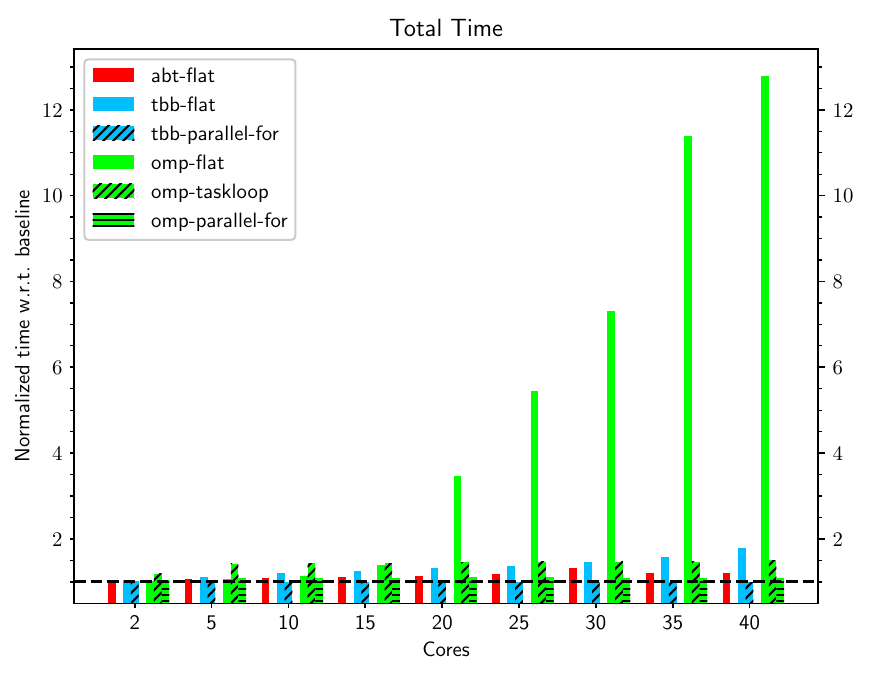}
	\includegraphics[width=0.49\linewidth]{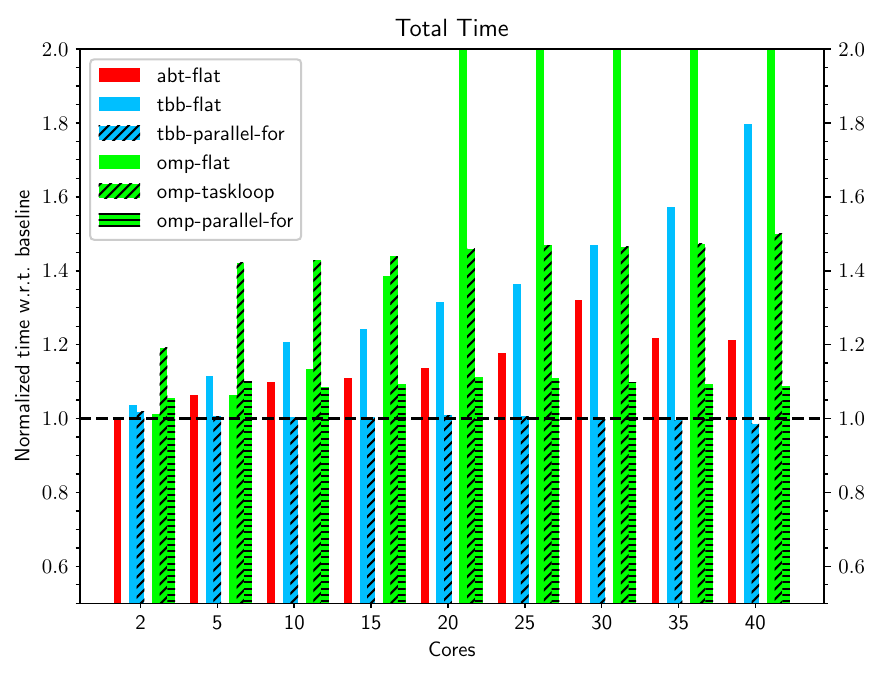}
	\caption{Left: Normalized total running time of high level constructs and 
	the \texttt{flat} model. Right: zoom-in at the range $0.5$-$2.0$.}
	\label{fig:high-level-vs-flat}
\end{figure}

\paragraph{Higher-level parallel constructs and the \texttt{flat} model:}
For the first benchmark, the \texttt{flat} tasking creation model is employed 
for each back-end and compared against higher-level 
constructs such as \ompfor,  \omptaskloop  and \tbbfor.
As expected, 
all back-ends exhibit an overhead when using the \texttt{flat} strategy compared 
to higher-level constructs  due to the cost of sequentially creating all the tasks.
Figure~\ref{fig:high-level-vs-flat} presents the running time normalized with respect 
to the baseline implementation. 
\texttt{omp-flat} back-end
suffers from the highest overhead, especially when 
more than $20$ cores are used,
 which is the size of the socket for this machine.
This trend is in part attributed to the fact that the naive creation of tasks in 
the \texttt{flat} model along with the \texttt{untied} specification allows any 
task to run on any core without any consideration about the affinity of
data with respect to the cores.
The higher level constructs perform better than their \texttt{-flat} counterparts. 
\omptaskloop improves significantly over \texttt{omp-flat} 
by merging multiple loop iterations into a single task, thus, decreasing the
number of tasks that need to be created and scheduled. Moreover, by creating and
scheduling fewer tasks, the number of context switches and cache-line
invalidations is also reduced.
\ompfor equipped with the \texttt{dynamic} 
scheduler performs even 
better thanks to the absence of the overhead of task creation.
\tbbfor outperforms the rest by dynamically adjusting the loop ranges
assigned to each task, based on the number of threads, 
the time for each task execution, and hardware occupancy.

\begin{figure}[!htpb]
	\centering
	\includegraphics[width=0.49\linewidth]{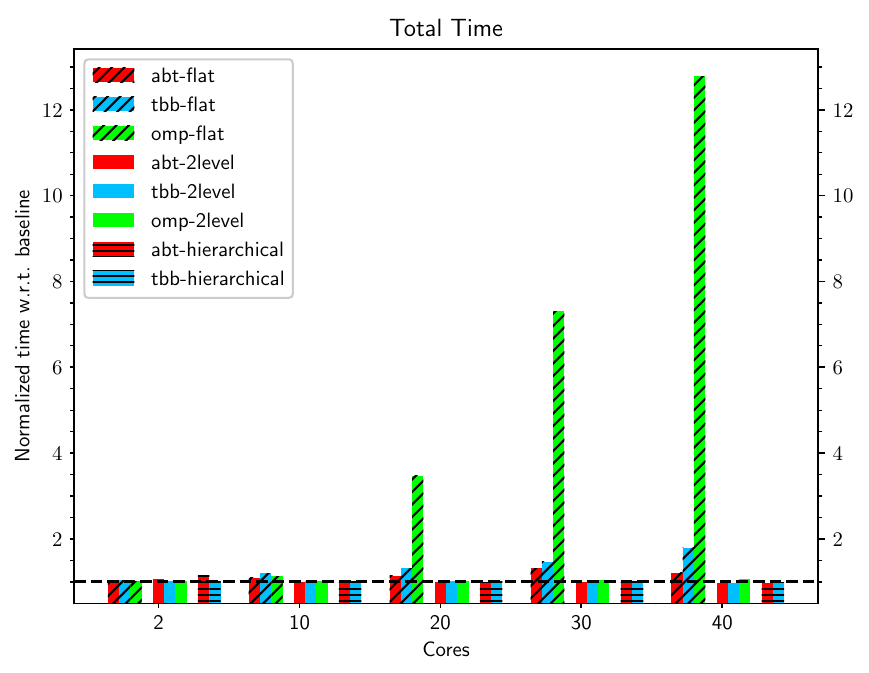}
	\includegraphics[width=0.49\linewidth]{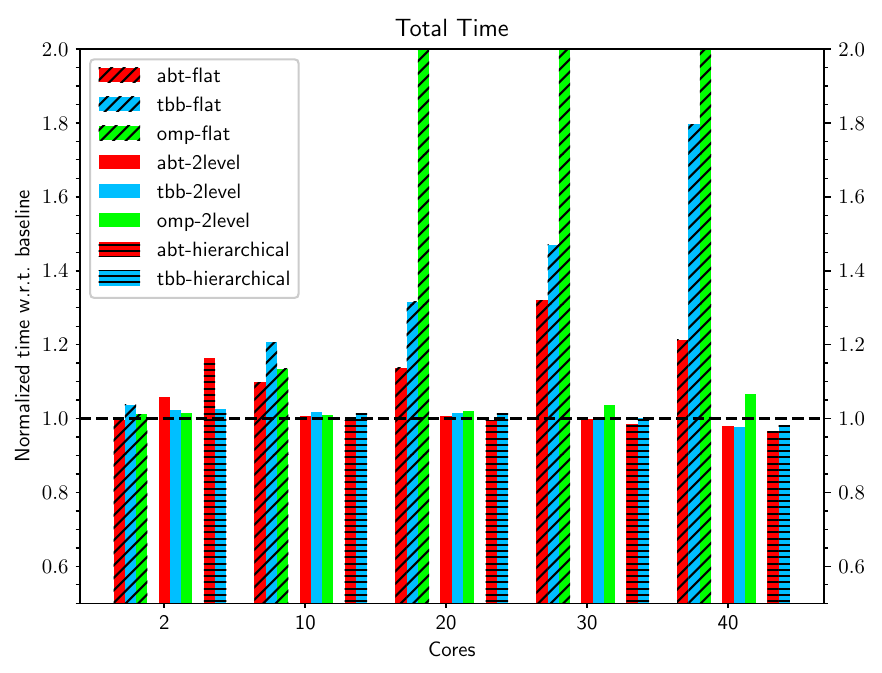}
	\caption{Left: Normalized total running of the three task creation 
	strategies implemented across the three different back-ends.
	The \texttt{grainsize} is fixed to $1$. 
	Right: zoom-in at the range $0.6$-$2.0$.}
	\label{fig:flat-vs-two-level-vs-hierarchical}
\end{figure}

\paragraph{Comparison between the \texttt{flat}, \texttt{2level} and \texttt{hierarchical} task creation strategies:}
In Figure~\ref{fig:flat-vs-two-level-vs-hierarchical}, the three task creation 
strategies are compared with each other.
Both the \texttt{2level} and the \texttt{hierarchical} strategies
reduce significantly the overhead in comparison to the 
\texttt{flat} strategy.
In the \texttt{2level} strategy, $2 \cdot nthreads$ level-1 tasks that 
partition the 
\texttt{user\_task\_args} vector are created sequentially. Then, each level-1 
task generates tasks that apply \emph{Operator} to the appropriate unit of 
work based on Table~\ref{tab:operation-characteristics}. For this dataset, the 
\texttt{grainsize} is set to $1$, which results in creating a level-2 task for 
each unit of work. 
Both the \texttt{2level} and the \texttt{hierarchical} strategies exhibit 
higher overhead at $2$  cores due to 
the fact that more tasks are created in total.
However, this overhead is 
amortized at a higher number of cores. The dual socket nature of the machine 
affects the system by a smaller amount, in comparison to the \texttt{flat}
strategy, with the \texttt{omp} back-end  
suffering from the highest overhead at about $7\%$ on $40$ cores. 
On the other hand, the \texttt{abt} and \texttt{tbb} back-ends achieve 
a small improvement when using $40$ cores.

The \texttt{hierarchical} task creation strategy creates tasks recursively by bisecting the 
\texttt{user\_task\_args} vector and creating two child tasks each time. The 
algorithm continues up until the target range reaches the grainsize, which is 1 
in this dataset. \texttt{abt-hierarchical} and 
\texttt{tbb-hierarchical} exhibit a 
higher overhead at $2$ cores, possibly due to the larger number of generated 
tasks. For more than $2$ cores, the \texttt{hierarchical} strategy
performs slightly better than the \texttt{2level}.
This is attributed, in part, to the fact that the
\texttt{hierarchical} strategy gives more flexibility in scheduling by having
many smaller tasks running concurrently (versus the \texttt{2level} which combines 
them in larger ones). This also creates more work-steal opportunities for 
idle threads, while at the same time avoids the overhead of creating tasks
sequentially (contrary to the  \texttt{flat} strategy). 
Results of applying the \texttt{hierarchical} strategy with a 
grainsize of $1$ using the \texttt{omp} back-end are omitted due to their high 
overhead, which reaches up to a $160$x slowdown on $40$ threads.  

\paragraph{Effect of \texttt{grainsize} for each task creation strategy:}
In the next dataset,
we demonstrate that the tasking framework in addition
contributes towards automating the process of performance tuning.
Since the tasking framework uses the same
scheduler across the four different operations of this case-study, 
running the application repeatedly while scanning through a 
set of different \texttt{grainsize} values and the available back-ends, 
we can obtain optimal values for each operation.
The grainsize
controls how many applications of \emph{Operator} will be bundled into a single task. 
In general, creating a high number of tasks (smaller grainsize) gives more 
flexibility for load balancing 
by the scheduler. However, a high number of small tasks increases the 
cost of load balancing. 
Previous studies on CDT3D~\cite{drakopoulos_fine_grained_2019} 
revealed  a significant dependence
of the running time on the number of buckets created during local 
reconnection. In
this study, instead of targeting a fixed number of buckets, we fix the size of each 
bucket to $150$ tetrahedra which was found to be ideal for the baseline application.

\newcommand\lw{0.5}
\begin{figure}[!htbp]
	\centering
	\includegraphics[width=\lw\linewidth]{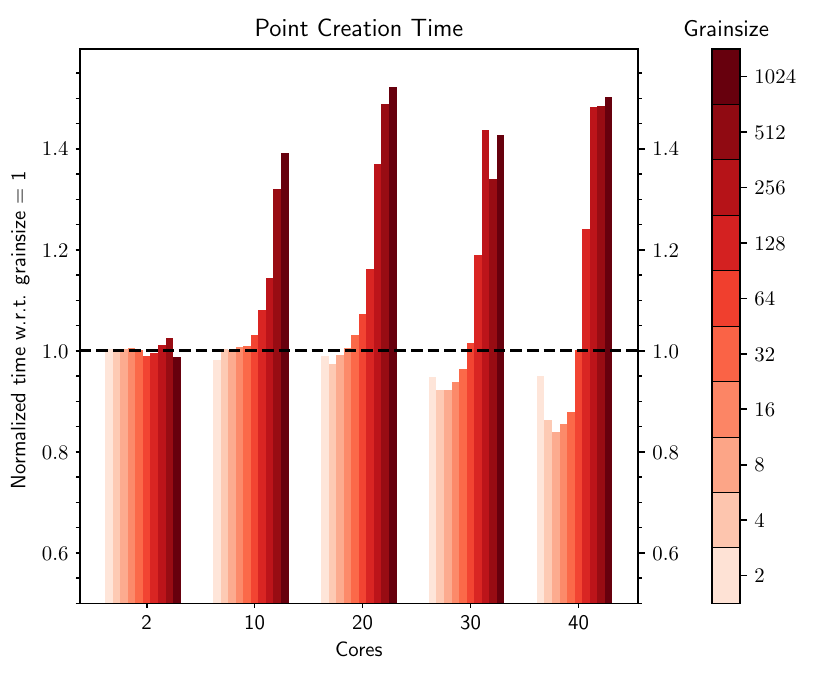}%
	\includegraphics[width=\lw\linewidth]{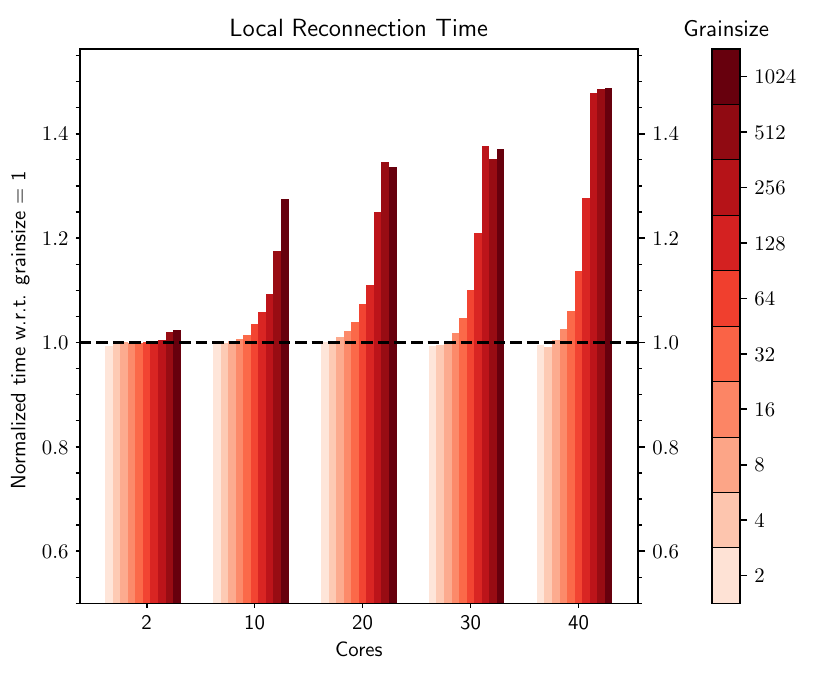}
	\includegraphics[width=\lw\linewidth]{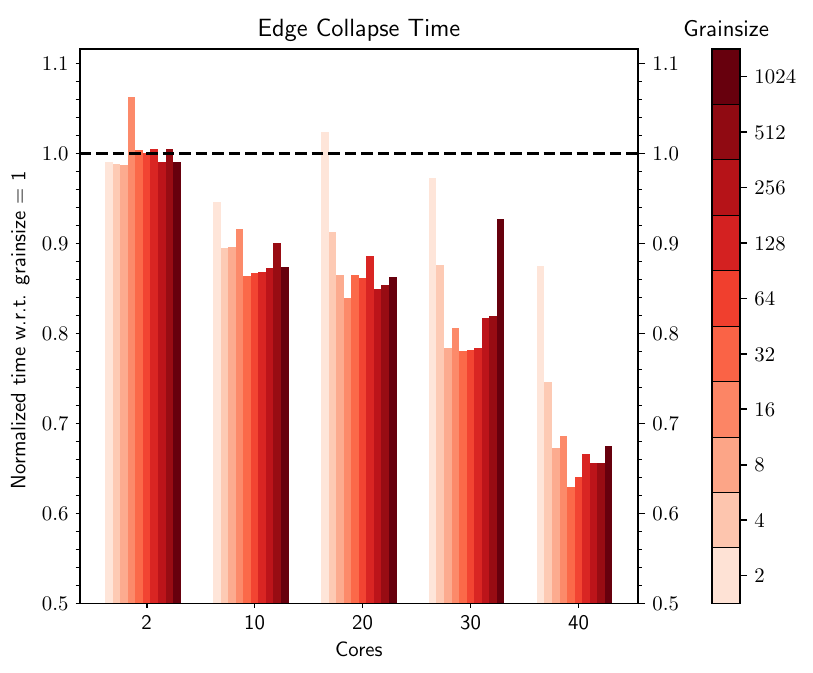}%
	\includegraphics[width=\lw\linewidth]{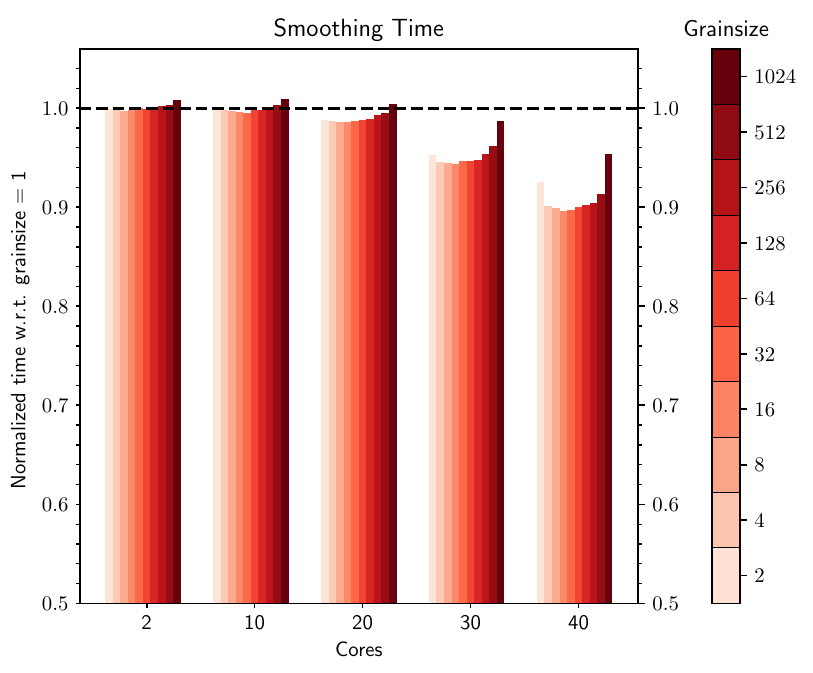}%
	\caption{Effect of grainsize for each operation for  \texttt{omp-2level}.
		Times are normalized based on the
		time taken using grainsize = 1.}
	\label{fig:grainsize-omp}
\end{figure}

\begin{figure}[!htbp]
\centering
	\includegraphics[width=\lw\linewidth]{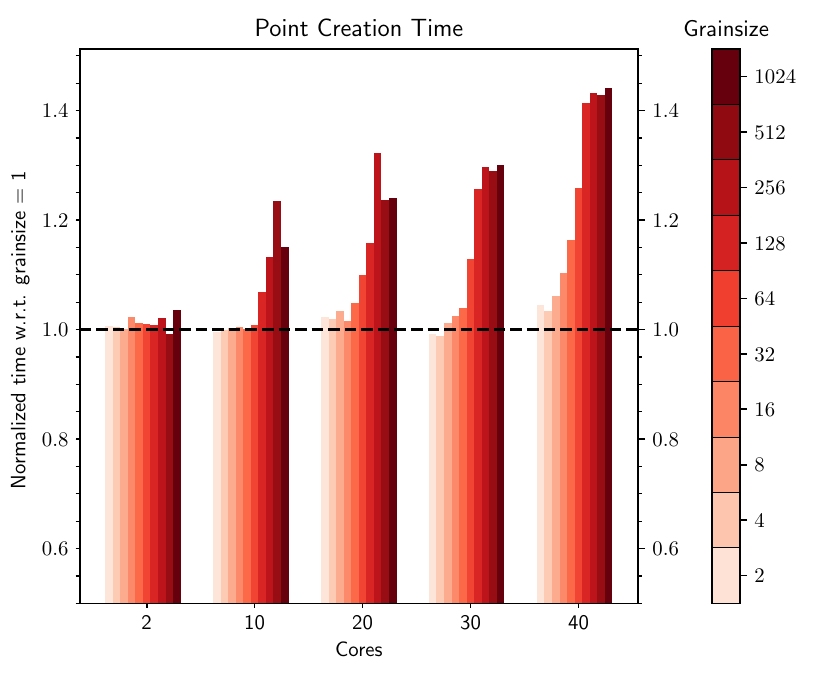}%
	\includegraphics[width=\lw\linewidth]{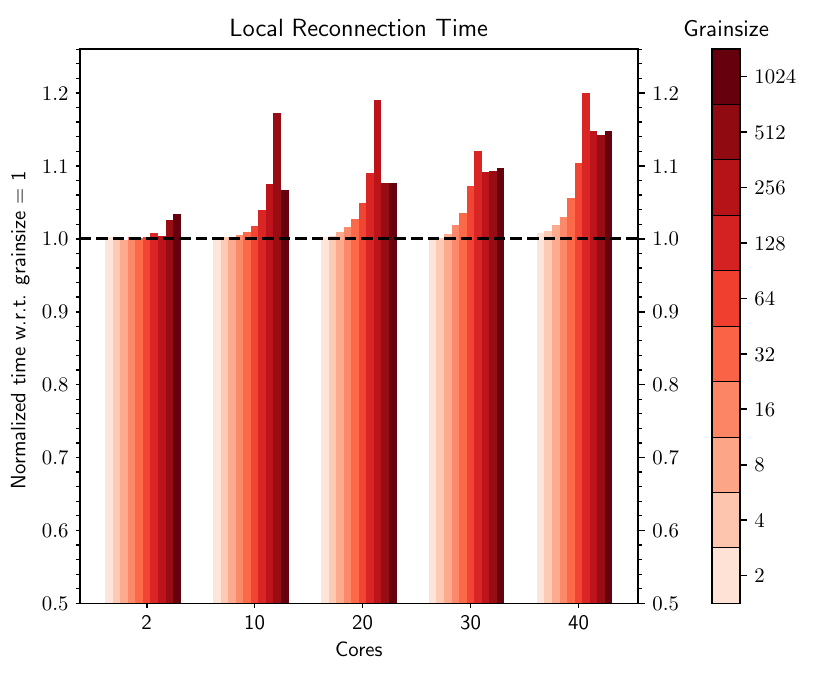}
	\includegraphics[width=\lw\linewidth]{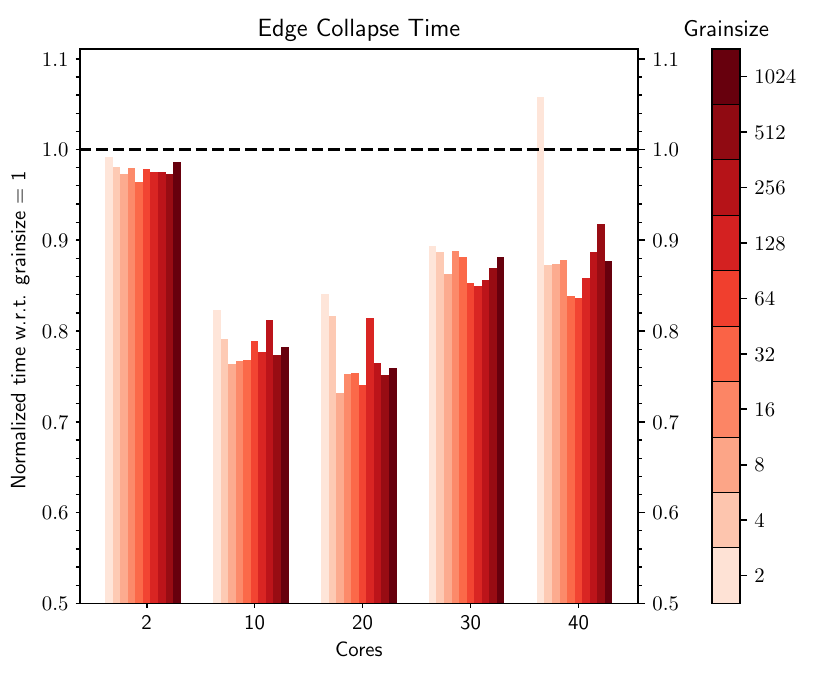}%
	\includegraphics[width=\lw\linewidth]{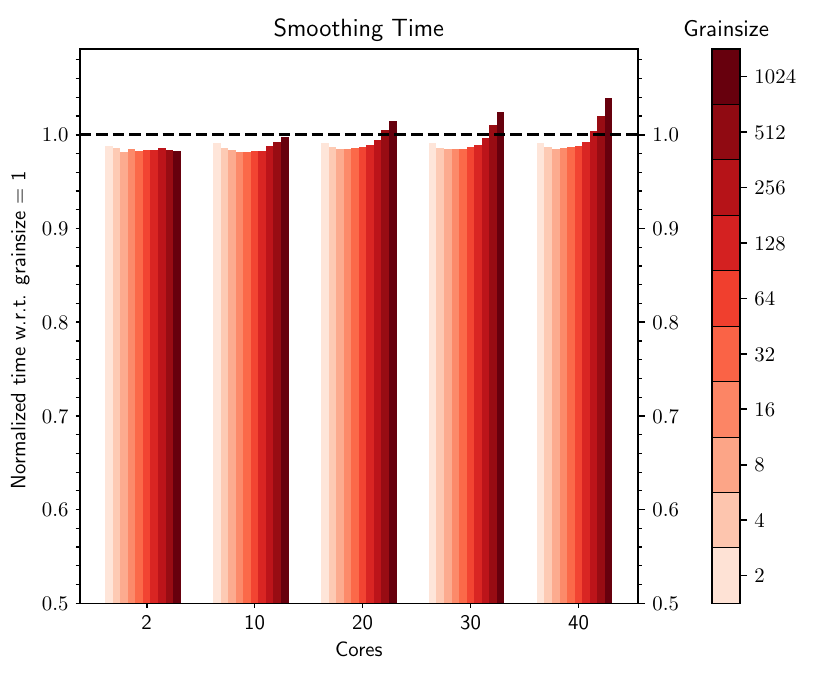}%
	\caption{Effect of grainsize for each operation for \texttt{tbb-2level}.
	Times are normalized based on the
	time taken using grainsize = 1.}
	\label{fig:grainsize-tbb}
\end{figure}

\begin{figure}[!htbp]
\centering
\includegraphics[width=\lw\linewidth]{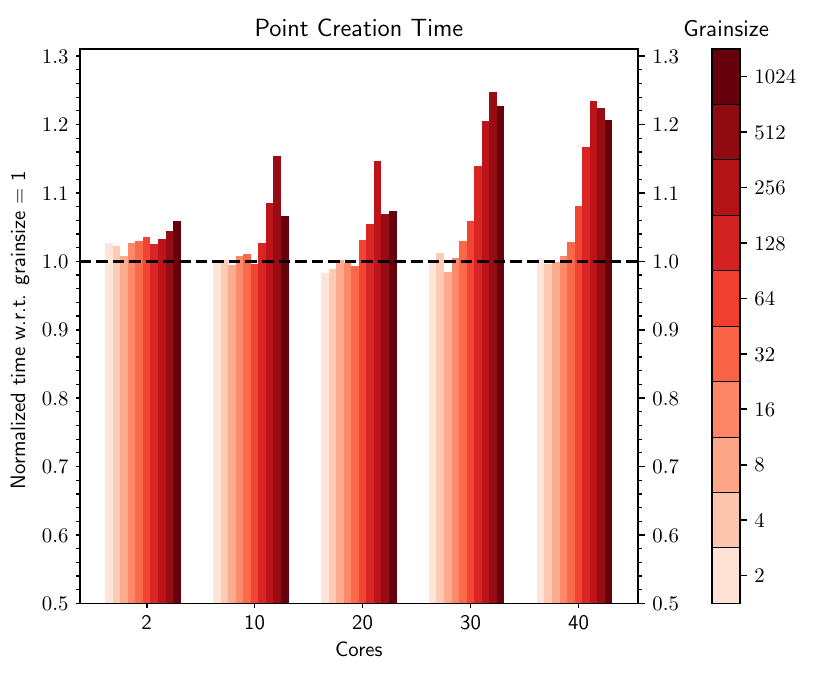}%
\includegraphics[width=\lw\linewidth]{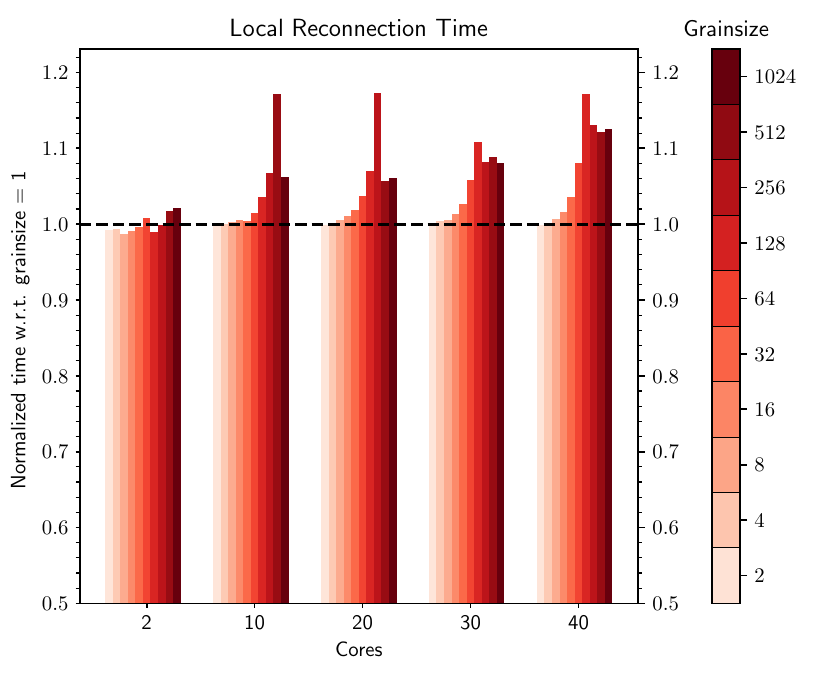}
\includegraphics[width=\lw\linewidth]{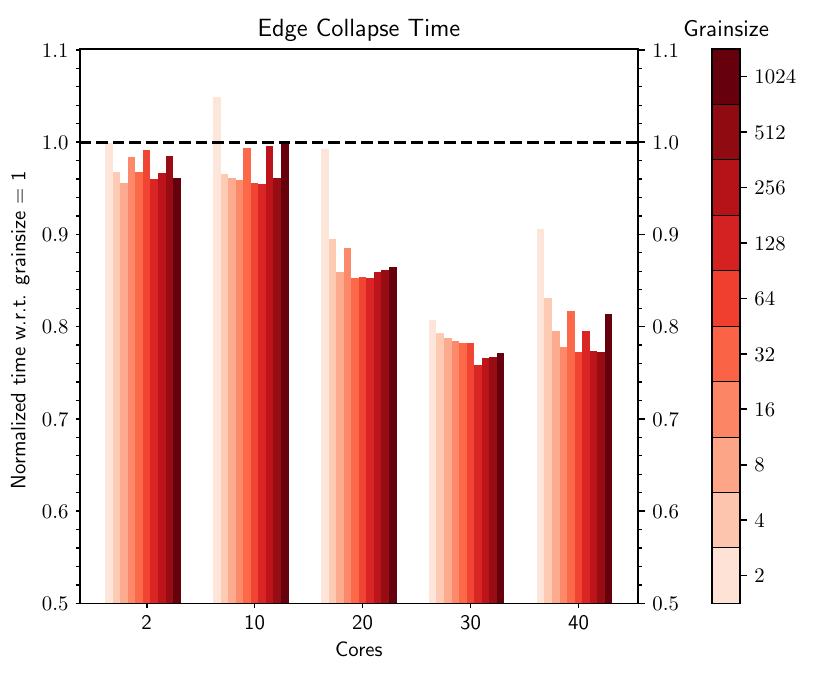}%
\includegraphics[width=\lw\linewidth]{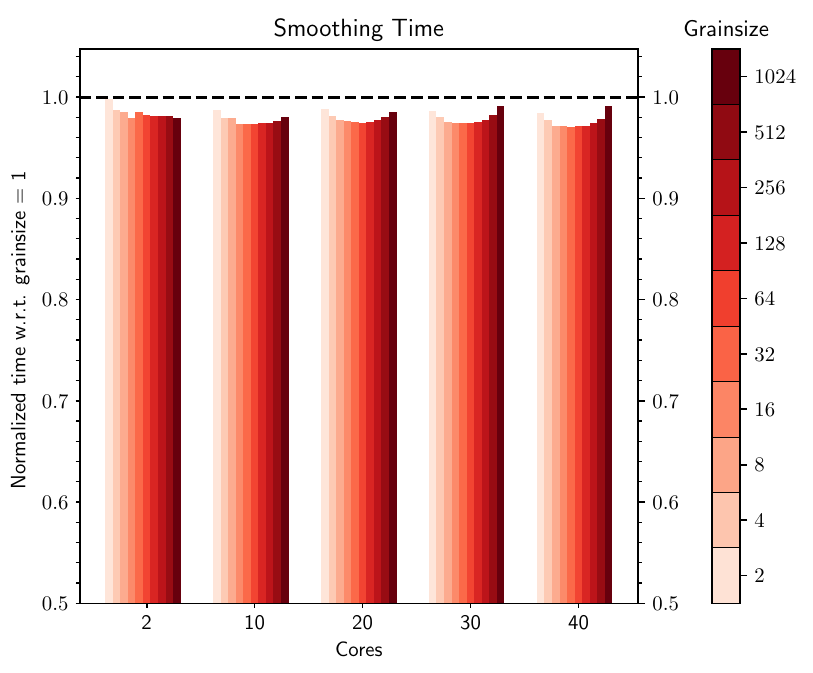}%
\caption{Effect of grainsize for each operation for \texttt{abt-2level}.
	Times are normalized based on the
	time taken using grainsize = 1.}
\label{fig:grainsize-abt}
\end{figure}

Figures~\ref{fig:grainsize-omp},~\ref{fig:grainsize-tbb} and~\ref{fig:grainsize-abt} 
compare the effect of different grainsizes for each operation
using the \texttt{2level} task 
creation strategy. 
The running time in each case is 
normalized based on the time achieved using a fixed grainsize of $1$. Overall, 
there are similar trends among the different back-ends. Point Creation and 
Local 
Reconnection perform better with a smaller grainsize. This is due to the fact
that these operations already decompose their data into ``buckets''
(see  Table~\ref{tab:operation-characteristics}) and each ``bucket'' offers enough 
workload to amortize 
the cost of creating and handling tasks. 
Using a higher grainsize creates fewer tasks, thus constraining the load balancer
and causes a loss in performance.
On the other hand, Edge Collapse and Vertex Smoothing, where the \emph{Operator} is 
designed to accept a single vertex, benefit significantly from increasing the 
grainsize. In particular, a grainsize of $128$  for the Edge Collapse offers 
more than $30\%$ speedup in 
comparison to a value of $1$ for the \texttt{omp} back-end and about $20\%$ for 
the other two back-ends. The gains for Vertex Smoothing are lower, but they also 
appear in the middle of the range which we experimented. 
The same analysis was also performed for the \texttt{hierarchical} strategy.
\texttt{abt-hierarchical} and \texttt{tbb-hierarchical} 
obtain optimal performance
for the same grainsize values, while for  \texttt{omp-hierarchical} the 
optimal values are 
$8192$ for Edge Collapse and Vertex Smoothing, $32$ for Vertex Creation
and $1$ for Local Reconnection. The graphs for the \texttt{hierarchical} 
strategy are  omitted for brevity. It should be noted that a different set of values 
could be ideal for a different configuration (hardware, meshing problem, etc.).


\newpage

\begin{figure}[!htpb]
\centering
\begin{subfigure}[t]{\textwidth}
\centering
\includegraphics[width=0.5\linewidth]{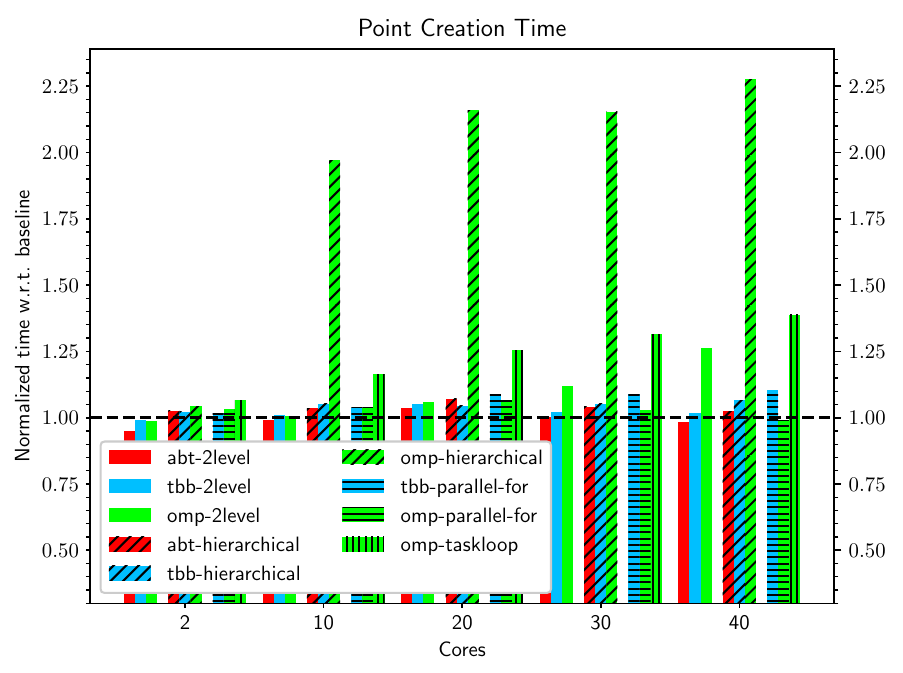}%
\includegraphics[width=0.5\linewidth]{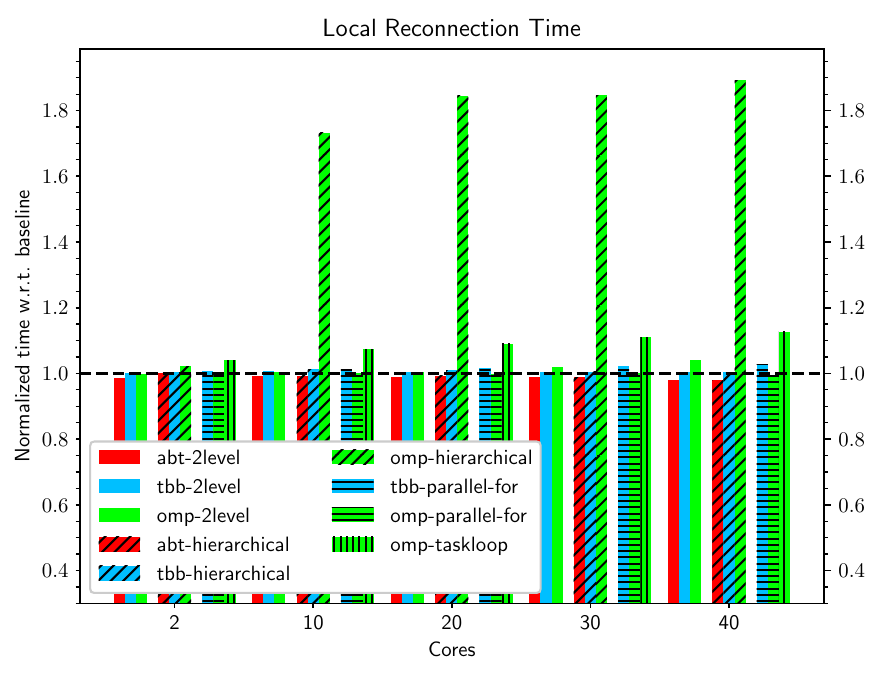}
\caption{Percent (\%) Improvement over \texttt{baseline} for the Vertex Creation and Local Reconnection.}
\end{subfigure}
\begin{subfigure}[t]{\textwidth}
\centering
\includegraphics[width=0.5\linewidth]{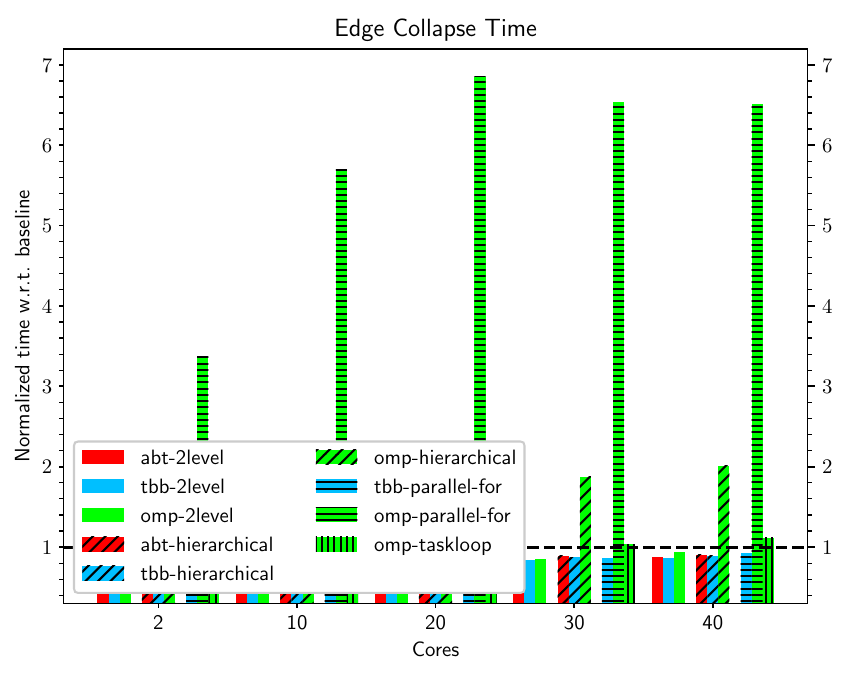}%
\includegraphics[width=0.5\linewidth]{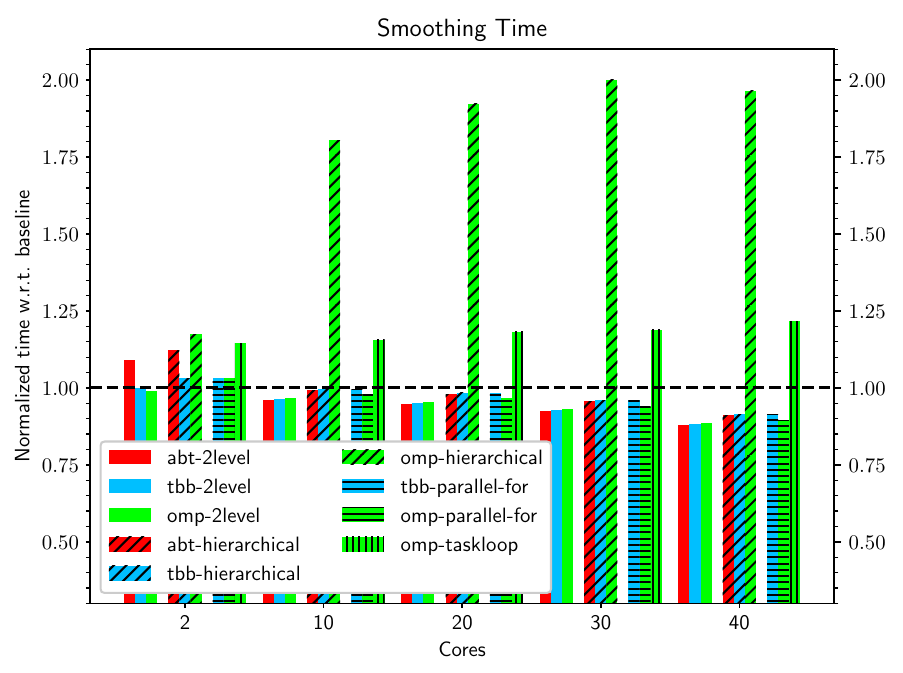}
\caption{Percent (\%) Improvement over \texttt{baseline} for the Edge Collapse and Smoothing Operations.}
\end{subfigure}
\caption{Performance improvements over the baseline implementation 
for the different back-ends using the \texttt{2level} and \texttt{hierarchical} 
strategies and optimal grainsizes. }
	\label{fig:opt}
\end{figure}

\paragraph{Performance of \texttt{2level} and \texttt{hierarchical} task creation strategies utilizing  optimal \texttt{grainsize}:}
Finally, we compare the  \texttt{2level} and \texttt{hierarchical}
task creation strategies utilizing the three different back-ends 
and the  optimal \texttt{grainsize} values derived in the previous paragraph.
The performance data indicate significant improvements for some back-ends
especially for the least optimized mesh operations (Edge Collapse, Vertex Smoothing).
Figure~\ref{fig:opt} depicts the  performance gains replacing the 
baseline implementation with the tasking framework for each of the operations. 
The grainsize is set to $1$ for the Vertex Creation and 
Local Reconnection, $128$ for Edge Collapse and $64$ for Vertex Smoothing 
when utilizing \texttt{*-2level}, \texttt{abt-hierarchical} 
or \texttt{tbb-hierachical}. For  \texttt{omp-hierarchical},
we used the grainsizes mentioned in the previous paragraph. 
Tables~\ref{tab:point-creation-local-reconnection-table}
and~\ref{tab:edge-collapse-smoothing} 
present the percent (\%) improvement over the baseline implementation.

\begin{table}[!h]
	\centering
		\caption{Percent $(\%)$ improvement of running time with respect to the baseline
		implementation for the  Point Creation and Local Reconnection operations. Negative numbers signify percent $(\%)$ slowdown.}
	\small
	\begin{tabular}{lrrr | rrr}
		& \multicolumn{3}{c}{Point Creation}              &  \multicolumn{3}{c}{Local Reconnection } \\
		& \multicolumn{3}{c}{Cores}   & \multicolumn{3}{c}{Cores} \\
		
		&  2      & 10       & 40      &  2      & 10       & 40      \\     \hline         
		abt-2level        & 5.03    & 0.82     & 1.47    & 1.45    & 0.79     & 2.01    \\
		tbb-2level        & 0.68    & -0.85    & -1.69   & -0.13   & -0.66    & -0.35   \\
		omp-2level        & 1.16    & -0.81    & -26.42  & 0.25    & -0.49    &  -4.11  \\
		abt-hierarchical  & 2.69    & -3.54    & -2.47   & -0.02   & 0.67     &  2.04   \\
		tbb-hierarchical  & -2.22   & -5.30    & -6.65   & -0.40   & -1.20    &  -0.32  \\
		omp-hierarchical  & -4.24   & -97.11   & -127.64 & -2.19   & -73.31   & -89.30  \\
		tbb-parallel-for  &  -1.58  &  -3.79   & -10.29  & -0.59   &   -1.07  &   -2.78 \\
		omp-taskloop      & -6.64   &  -16.30  &  -38.80 & -4.07   & -7.35    & -12.67  \\
		omp-parallel-for  &  -3.39  &  -3.65   &   1.01  & -0.22   &  -0.27   &   0.53  \\
	\end{tabular}
	\label{tab:point-creation-local-reconnection-table}
\end{table}
\begin{table}[!h]
	\centering
		\caption{Percent $(\%)$ improvement of running time with respect to the baseline implementation of the Edge Collapse and Smoothing operations.
		Negative numbers signify percent $(\%)$ slowdown.}
	\small
	\begin{tabular}{l rrr | rrr}
		& \multicolumn{3}{c}{Edge Collapse}  & \multicolumn{3}{c}{Smoothing } \\
		& \multicolumn{3}{c}{Cores}          & \multicolumn{3}{c}{Cores}             \\
		&  2      & 10       & 40            &  2      & 10      & 40     \\ 
		\hline                                                                                
		abt-2level        & -0.46   & 10.88    & 12.05         & -9.18   & 3.98    & 12.04  \\
		tbb-2level        & 7.74    & 10.60    & 13.42         & -0.07   & 3.68    & 11.60  \\
		omp-2level        & 8.69    & 8.87     &  5.88         & 0.98    & 3.26    & 11.45  \\
		abt-hierarchical  & -5.44   & 5.75     &  9.67         & -12.32  & 0.77    & 8.92   \\
		tbb-hierarchical  & 3.02    & 5.26     &  10.58        & -3.17   & 0.49    & 8.62   \\
		omp-hierarchical  & -13.43  & -76.16   & -101.07       & -17.42  & -80.47  & -96.60 \\
		tbb-parallel-for  & 2.28    &  3.86    &   6.86        & -3.31   & 0.54    &   8.49 \\
		omp-taskloop      & -6.53   & -8.32    & -11.44        & -14.47  & -15.59  & -21.59 \\
		omp-parallel-for  & -238.23 &  -469.65 & -550.96       & -3.16   &  1.89   & 10.67
	\end{tabular}
	\label{tab:edge-collapse-smoothing}
\end{table}

The Point Creation and Local Reconnection operations benefit the least. 
The difference in performance gains between the two pairs of
operations is related to the fact that not 
all operations use the same back-end in the baseline implementation (see
Table~\ref{tab:operation-characteristics}).  In particular, the Point Creation
and Local Reconnection operations utilize a custom work-sharing approach described
in~\cite{drakopoulos_fine_grained_2019} which has been 
optimized for the these operations. On the other hand, the Edge Collapse and 
Vertex Smoothing operation were parallelized using simple OpenMP primitives.
Also, the first two operations operate on ``buckets'' (i.e., lists of elements) 
instead of single elements, thus introducing \emph{a-priori} data decomposition which
may limit the effect of using different scheduling techniques.
 
\texttt{abt-2level}  performs the best, 
offering up to $1.47\%$ and $2.01\%$ improvement on $40$ cores for the Point Creation 
and Local Reconnection operations, respectively. The Edge Collapse 
and Smoothing operations benefit more. \texttt{tbb\-2level} performs the best for the 
Edge Collapse operation delivering more than $13\%$ improvement on $40$ cores, while for
Smoothing, the best performing is \texttt{abt-2level} with up to $12\%$
improvement over the baseline implementation.
For comparison, we also append data from the higher-level
constructs
(\tbbfor, \ompfor, \omptaskloop) 
(i.e., from Figure~\ref{fig:high-level-vs-flat})
which should serve as a reference point, since they provide the simplest
way to introduce tasks within an application.
The higher-level constructs fail to improve
the performance for the operations that use custom scheduling, and only some
of them deliver small gains for Edge Collapse and Vertex Smoothing. 
In particular, \tbbfor delivers improvements 
for Edge Collapse and Smoothing and \ompfor exhibits some gains for Smoothing.
However, the gains using the same back-ends within the 
proposed approach are higher.

\begin{figure}[!htpb]
	\centering
	\includegraphics[width=0.7\linewidth]{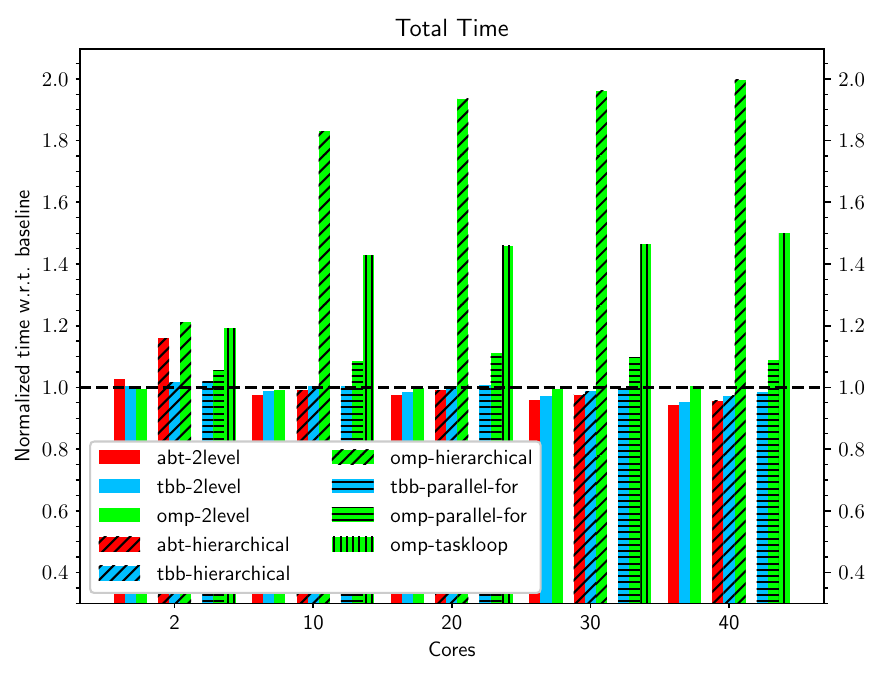}
	\caption{Total running time of the entire application with optimal grainsizes for each
	back-end and task creation strategy.}
	\label{fig:opt-total}
\end{figure}

Figure~\ref{fig:opt-total} and Table~\ref{tab:total-time} depict 
the effect of the tasking framework on the total running time of the application while utilizing
the optimal grainsize for each back-end and task creation strategy.
Overall, \texttt{abt-2level}
performs the best with up to $5.81\%$ 
improvement  on $40$ cores.
\texttt{tbb-2level} offers a slightly smaller
improvement $(4.72\%)$ while 
\texttt{omp-2level} adds a small overhead $(-0.52\%)$ on $40$ cores. 
Although, the \texttt{hierarchical} strategy is able to exploit concurrency at an 
earlier stage, it does not perform as well as the \texttt{2level} strategy. 
This is attributed, in part, to the fact that the \texttt{2level} strategy generates
almost half the number of tasks in comparison to the 
\texttt{hierarchical} strategy.
In particular, the \texttt{2level} strategy  generates
$2 \cdot nthreads$ $+ n/grainsize$ 
tasks while the hierarchical generates 
$2^{\log_2(n/grainsize) +1 } -1 = 2(n/grainsize) -1$ tasks
where, $n$ number of work-units passed to \texttt{task\_for} (i.e.,
the length of vector \texttt{user\_task\_args} in Listing~\ref{code:task_for})%
\footnote{$h=\log_2(n/grainsize)$ is the depth of a perfect binary tree with 
$n/grainsize$ terminal nodes. $2^{h+1} -1$ is the number of nodes for a  
perfect binary tree with depth $h$.}.
Since, in general, $2\cdot nthreads \ll n$ and $grainsize \ll n$, 
the \texttt{2level}
strategy produces about half the number of tasks.

\begin{table}
    \centering
        \caption{Percent (\%) improvement of total running time with respect to the baseline implementation.
    	Negative numbers signify percent $(\%)$ slowdown.}
    \begin{tabular}{l rrr}
                      & \multicolumn{3}{c}{Total Time}\\
                      & \multicolumn{3}{c}{Cores}  \\
                      &  2      & 10       & 40     \\ 
    \hline
    abt-flat          &  0.41   & -9.80   & -21.34 \\
    tbb-flat          &  -3.76  & -20.62  & -79.67 \\
    omp-flat          & -1.21  & -13.46  & -1177.87 \\\\
    
    abt-2level        & -2.81   & 2.31    & 5.81   \\
    tbb-2level        &-0.31    & 1.25    & 4.72   \\
    omp-2level        & 0.62    & 0.91    & -0.52  \\\\
    
    abt-hierarchical  & -15.86  & 0.91    & 4.39   \\
    tbb-hierarchical  & -1.62   & -0.31   &  2.79  \\
    omp-hierarchical  & -21.0   & -83.01  & -99.71 \\\\
    
	tbb-parallel-for  & -1.81   &  -0.30  &  1.40  \\
	omp-taskloop      & -19.17  & -42.88  &  -49.97  \\
	omp-parallel-for  & -5.48   &  -8.57  &  -8.73  \\
    \end{tabular}
    \label{tab:total-time}
    \end{table}

\newpage

\subsubsection{Stability of the Tasking Approach}
\label{sec:cdt3d-stability}
\begin{figure}[!htpb]
	\centering
	\begin{subfigure}[t]{0.5\textwidth}
	\includegraphics[width=\linewidth]{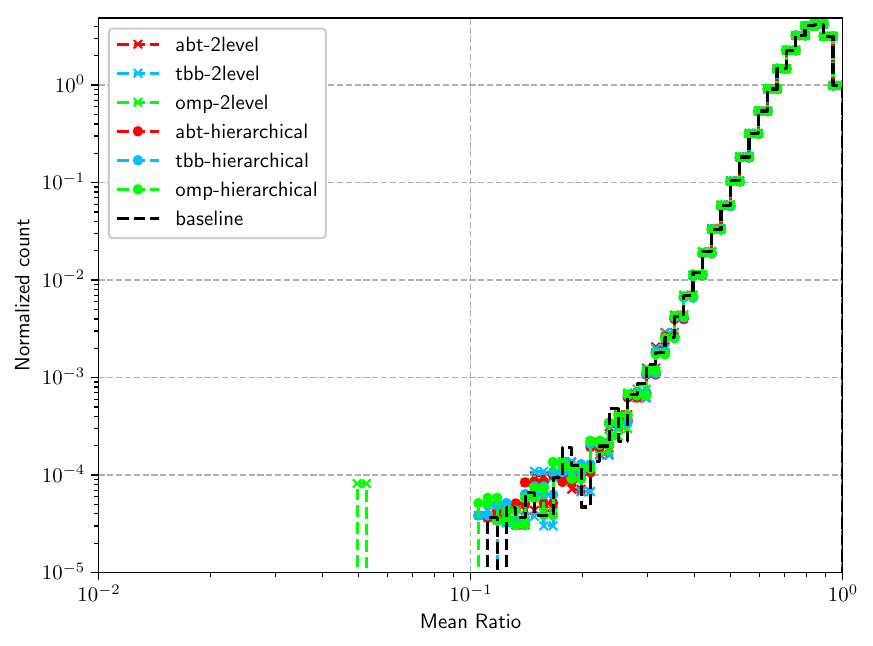}%
    \caption{}
	\label{fig:cdt3d-quality}
	\end{subfigure}%
	\hfill
	\begin{subfigure}[t]{0.5\textwidth}
	\includegraphics[width=\linewidth]{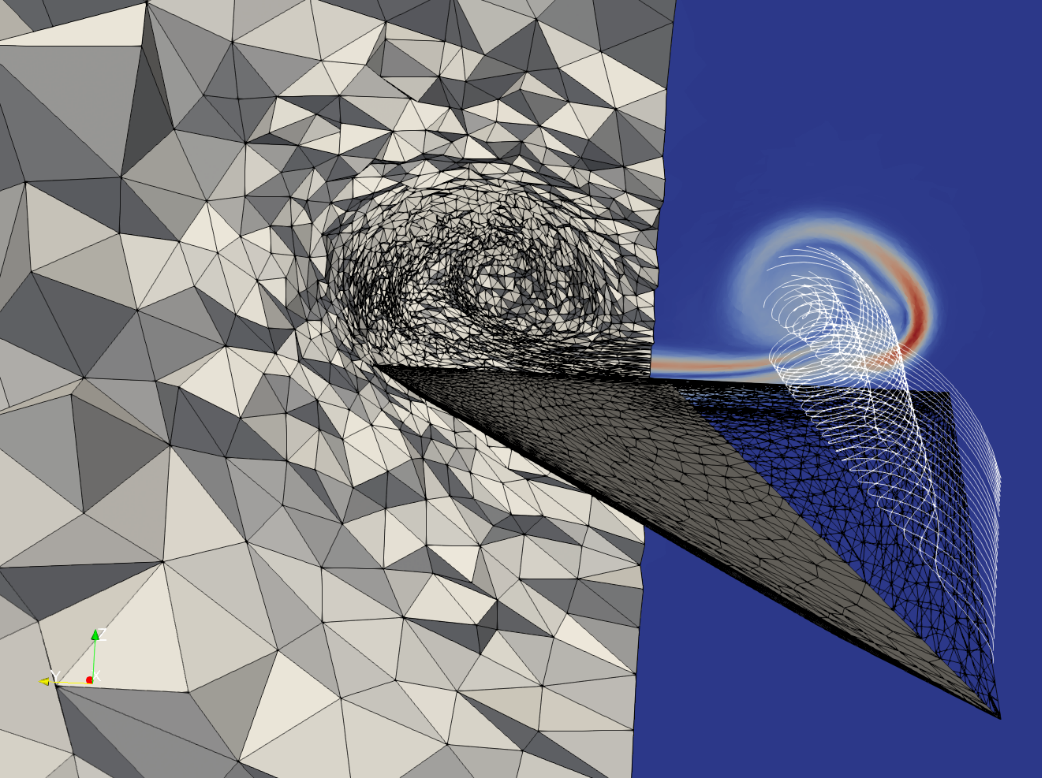}
    \caption{}
	\label{fig:cdt3d-delta-wing}
	\end{subfigure}
	\caption{Stability data and visualization of the generated mesh for this use-case.
	(a): Comparison of the mean-ratio quality metric for 
	the different back-ends.
	(b): Visualization of the mesh generated by the experiments in this section:
	Metric-adapted mesh to a laminar flow over a delta wing.}
	\label{fig:cdt3d-stability}
\end{figure}

Among the requirements for a parallel mesh generation code 
as presented in \cite{tsolakis_parallel_2021} is the one of \emph{stability}
which requires that a mesh generated in parallel has comparable quality 
with one generated sequentially by the same application. The stability of the baseline application
has been already demonstrated in~\cite{tsolakis_parallel_2021}. In 
Figure~\ref{fig:cdt3d-quality}, we compare a mesh quality measure 
among the different 
back-ends and task creation strategies
of the previous section.
In particular,
the histograms are built using the meshes generated at $40$ cores in 
Figure~\ref{fig:opt} and averaging the data over the $10$ runs of the experiment. Even 
when using a logarithmic scale, there is no significant difference between 
the different back-ends with the exception of the \texttt{omp-2level} 
back-end that produced a slightly lower minimum value. Still, the 
results are within the range ($>0.01$) produced 
by other state-of-the-art approaches as
presented in~\cite{tsolakis_parallel_2021}.

\newpage
\subsection{Case Study II: Parallel Optimistic Delaunay Meshing (PODM)}\label{sec:case-study-PODM}
The Parallel Optimistic Delaunay Meshing (PODM) method presented 
in~\cite{foteinos_high_2014} delivers good parallel performance on DSM machines 
and high mesh quality along with provable fidelity guarantees. In terms of
meshing operations, PODM initializes the meshing procedure with 
only  $6$ elements that decompose the bounding box of the input image. The mesh 
is incrementally refined by inserting points generated based on rules that 
guarantee the quality and fidelity of the mesh with respect to the input image.
For more details, see Figure~\ref{fig:podm_code}. The point insertion 
procedure is  built around  the Bowyer-Watson 
kernel~\cite{bowyer_computing_1981,watson_computing_1981} which introduces 
new points in the mesh while simultaneously preserving the invariant that after each point insertion, the mesh retains the  Delaunay property. 
There are many ways to decompose the Bowyer-Watson kernel 
into tasks.
 In the past, it has been decomposed into \emph{compute data 
dependencies (cavity)}, \emph{collect data dependencies} and \emph{update 
connectivity} tasks~\cite{chrisochoides_task_1996}. In higher dimensions 
($>3$), 
it is advantageous to decompose the data dependency evaluation (i.e., cavity 
expansion) into many tasks~\cite{foteinos_4d_2014}.
Other approaches~\cite{foteinos_dynamic_2011,marot_one_billion_2019}, 
transform the problem of Delaunay Mesh Refinement into two tasks: one of generating the vertices to be added and one that updates the current 
triangulation by inserting the vertices. In this study, in an effort to keep the problem complexity low and introduce 
only a small amount of code changes, only two types of tasks will be used; 
one for scheduling an element and one for refining it.

PODM caries many years of optimization for DSM 
machines~\cite{foteinos_phd_thesis_2013}.
However, as it happens with highly optimized codes, viewing
them from a new perspective may reveal new challenges. The optimizations 
and design decisions that made PODM very efficient
put constraints on the tasking implementation. The most
important one, is that threading is managed explicitly by the application and
the Load Balancing section of Figure~\ref{fig:podm_code} 
is responsible for populating the
work-queue of each thread. In other words, the workload distribution is 
explicit and tightly
integrated with the application.

To overcome this issue, we use the \texttt{thread\_id}
obtained by the threading environment in order to access the appropriate queue 
in a  thread-safe manner. Listing~\ref{code:podm_high_level_tasking_two} 
presents a  high level pseudocode
of the tasking version of PODM. It implements the \texttt{flat} model of 
Figure~\ref{fig:task_specturm} by decomposing the algorithm of
Figure~\ref{fig:podm_code}
into two tasks. \texttt{ScheduleTask} creates tasks for  a number of 
elements from a thread queue. Notice that the task created in
line~\ref{codeline:schedule_el} of Listing~\ref{code:podm_high_level_tasking_two}.
can run with any \texttt{thread\_id} which implicitly enables work 
distribution
between different threads.
Moreover, each thread will push the newly created elements
into its private queue in line~\ref{codeline:local_push} of Listing~\ref{code:podm_high_level_tasking_two}.
\texttt{RefineBadElement} encapsulates the blue section of
Figure~\ref{fig:podm_code} and it will generate the point to be inserted,
calculate and lock its cavity (i.e., data dependencies) and apply
the Bowyer-Watson kernel as well as release any acquired
locks in the end. 

\begin{figure}[!htpb]
		\centering
		\includegraphics[width=0.6\linewidth]{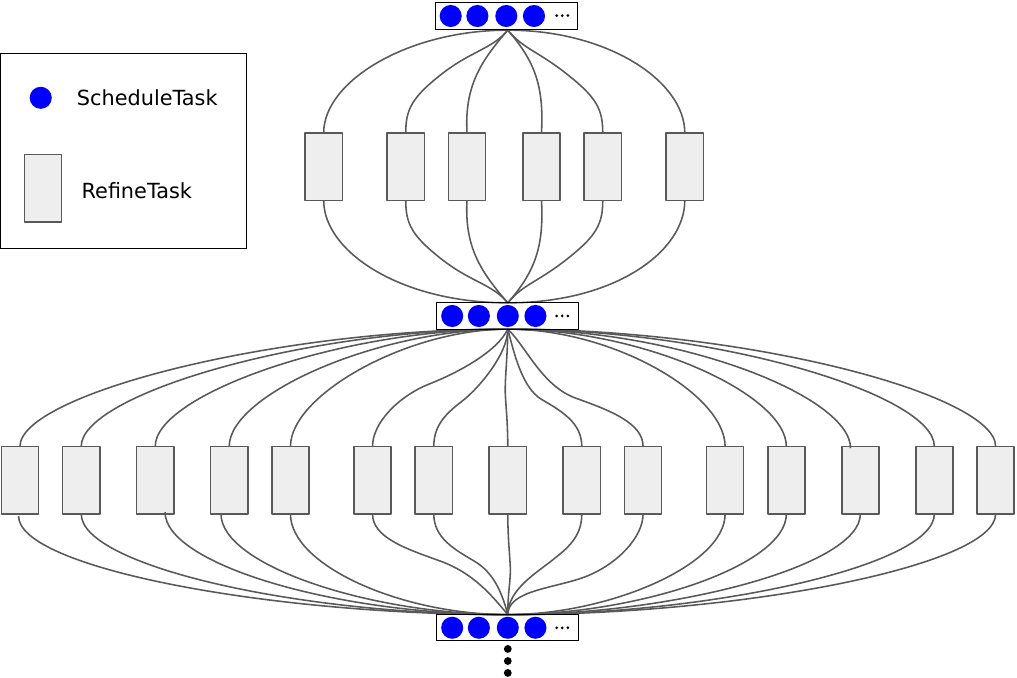}
		\captionof{figure}{Flowchart of the tasking version of PODM.}
		\label{fig:podm-tasks-flowchart}
\end{figure}

Figure~\ref{fig:podm-tasks-flowchart} depicts a high level view of the execution
flow of the tasking version. Initially,  \texttt{ScheduleTask} will spawn a 
task for each of the $6$ elements of the initial mesh. Since the initial mesh
is very small, only some of the initial $6$ tasks will be completed 
successfully due to rollbacks.
In the second round,  \texttt{ScheduleTask} will create a 
task for each of the newly created elements,  and the  process continues until all 
\texttt{thread\_Queue}s are empty. 
Notice  that this simple implementation has two major issues:
Possibility of livelocks, occurring when two tasks lock themselves in an
infinite cycle trying to acquire different parts of overlapping cavities, and
the algorithm termination depending on empty thread-local queues. Thread-local
queues are accessed based on the \texttt{thread\_id} acquired from the tasking environment,
which is, in general, random. Thus, there is a possibility that a  non-empty 
\texttt{thread\_Queue} may never get accessed.   
In these experiments, we didn't notice any of the aforementioned issues, 
but there is still a chance that they might occur.
In a follow-up study, we could 
integrate our previous work on contention managers~\cite{foteinos_high_2014} that can treat both issues efficiently.

Line~\ref{codeline:schedule_limit} of Listing~\ref{code:podm_high_level_tasking_two} includes a limit on the number of elements 
to be scheduled at a time. This is necessary since many of the generated tasks will
be invalid by the time they run because their corresponding element will have
been deleted as part of an operation executed on another cavity. 
Therefore, scheduling all available elements at once will generate a high number of
aborted tasks.

\begin{lstlisting}[
style=my_cpp_style,
caption={High level tasking-based pseudocode of PODM.},
label=code:podm_high_level_tasking_two
]
main()
{
  while(not all thread_Queues are empty){
    // Launch enough ScheduleTasks to keep all cores
    // busy
    for(tid : thread_ids) 
	    task::create_and_schedule_task(ScheduleTask);
	task::wait_for_all();
	}	
}
ScheduleTask()
{	
	int tid = task::get_thread_id();
	int scheduled = 0;
	while(scheduled < schedule_limit && 
	!(thread_Queue[tid].empty()))|\label{codeline:schedule_limit}|
	{
		el = thread_Queue[tid].pop();
		task::create_and_schedule_task(el,RefineTask);|\label{codeline:schedule_el}|
		scheduled++;		
	}
}
		
RefineTask(el)
{
	int tid = task::get_thread_id();
	success = el.lock_vertices()
	if(success)	{
		RefineBadElement(el);
		for(el : newly_created_elements) 
		  thread_Queue[tid].push(el)	|\label{codeline:local_push}|
	}
}
\end{lstlisting}

\subsubsection{Performance Evaluation}
\begin{figure}[!htpb]
	\centering
	\includegraphics[width=0.49\linewidth]{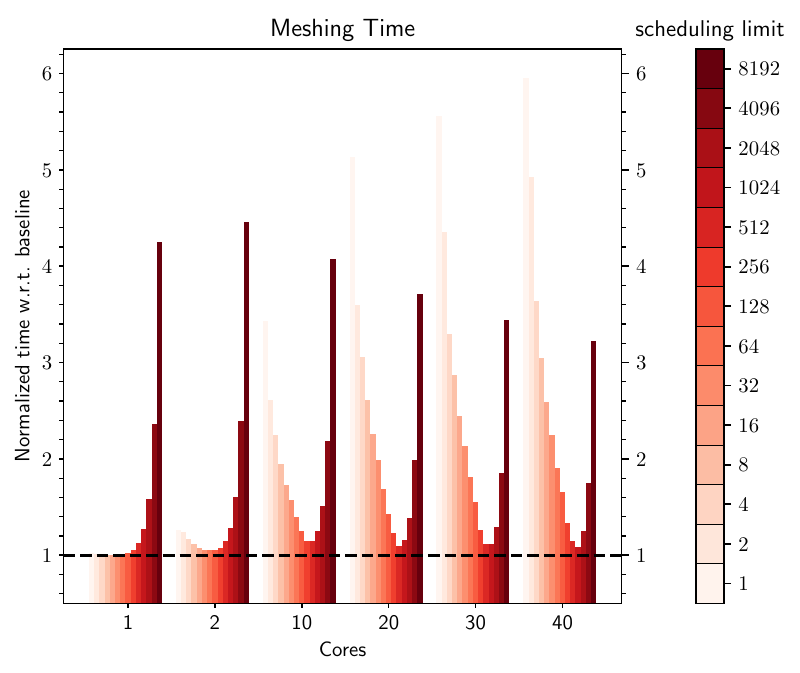}
	\includegraphics[width=0.49\linewidth]{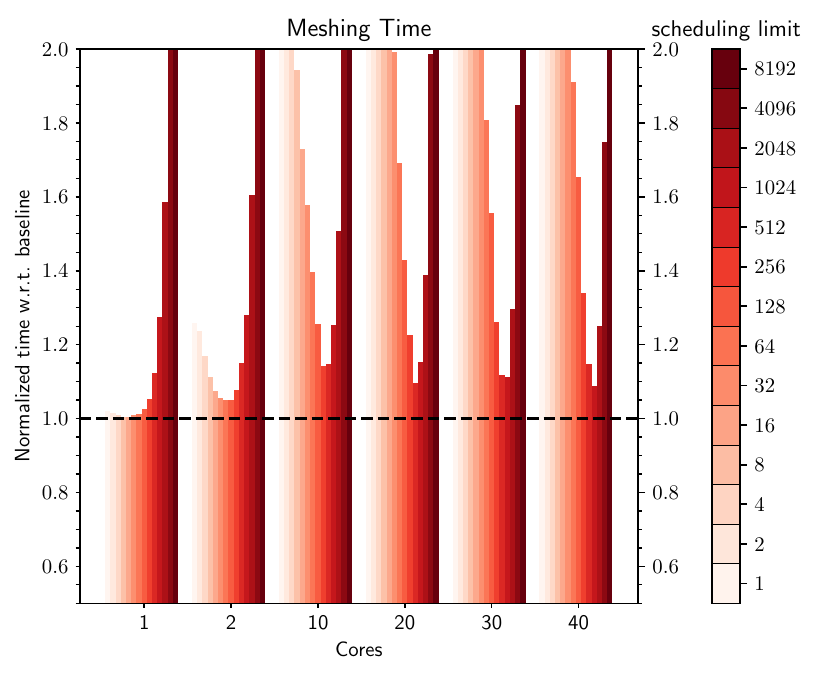}
	\caption{Effect of scheduling limit for the second approach using the \texttt{tbb}
	back-end. Right zoom-in in range 0.5-2.0}
	\label{fig:podm-tbb-schedule-limit}
\end{figure}

The hardware and compiler configuration is the same as in Section~\ref{sec:cdt3d-evaluation}.
Figure~\ref{fig:podm-tbb-schedule-limit} depicts the effect of the different
values of \texttt{schedule\_limit} to the runtime with respect to the 
baseline application.
Notice that for $1$ thread, the 
ideal value is low. Any limit below $128$ performs equally well,
while for $40$ a value of $512$ performs better since it provides the system 
with more concurrency, albeit at the expense of more aborted tasks. 

\begin{figure}[!htpb]
	\centering
	\includegraphics[width=0.7\linewidth]{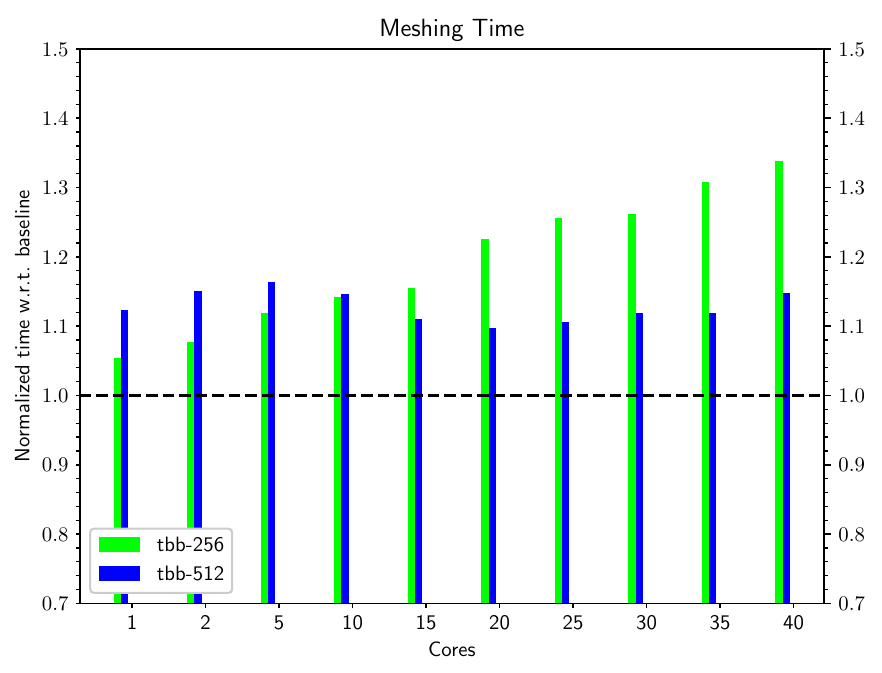}
	\caption{Normalized meshing time of the tasking version of PODM for two 
	different values of the \texttt{schedule\_limit}.}
	\label{fig:podm-tasks}
\end{figure}

Finally, Figure~\ref{fig:podm-tasks} presents the best values among our 
experiments. The use of the tasking framework adds only between $ 5\%-14\%$ overhead with 
respect to the highly optimized baseline application across the
different number of cores. 
Of course, these results come with the shortcomings mentioned above; but based on our previous experience, resolving them should not negatively impact  
the performance. 

The \texttt{abt} and \texttt{omp} back-ends exhibit much higher overheads in 
this case. Scheduling decisions and the internal optimizations of \texttt{tbb} 
could be one of the reasons. Investigating the cause of this overhead  and 
optimizing the 
\texttt{abt} and \texttt{omp} back-end implementations of the generalized 
framework could be investigated in the future.

\subsubsection{Stability of the Tasking Approach}
\label{sec:podm-stability}
\begin{figure}[!htpb]
	\centering
	\begin{subfigure}[t]{0.5\textwidth}
		\includegraphics[width=\linewidth]{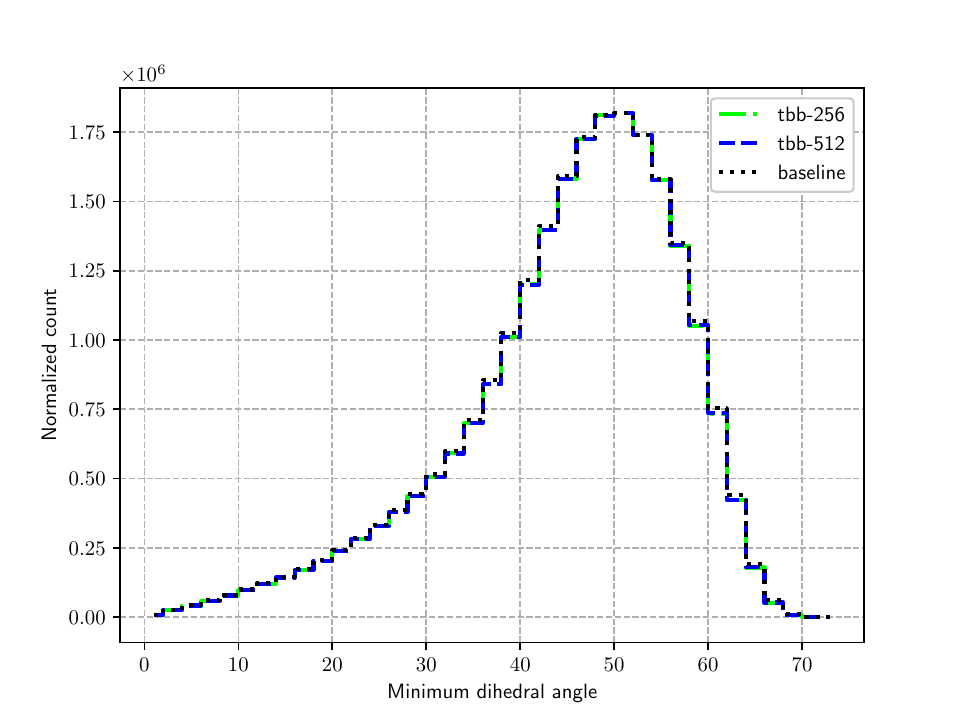}
		\caption{}
		\label{fig:podm-quality}
	\end{subfigure}%
	\begin{subfigure}[t]{0.5\textwidth}
		\includegraphics[width=\linewidth]{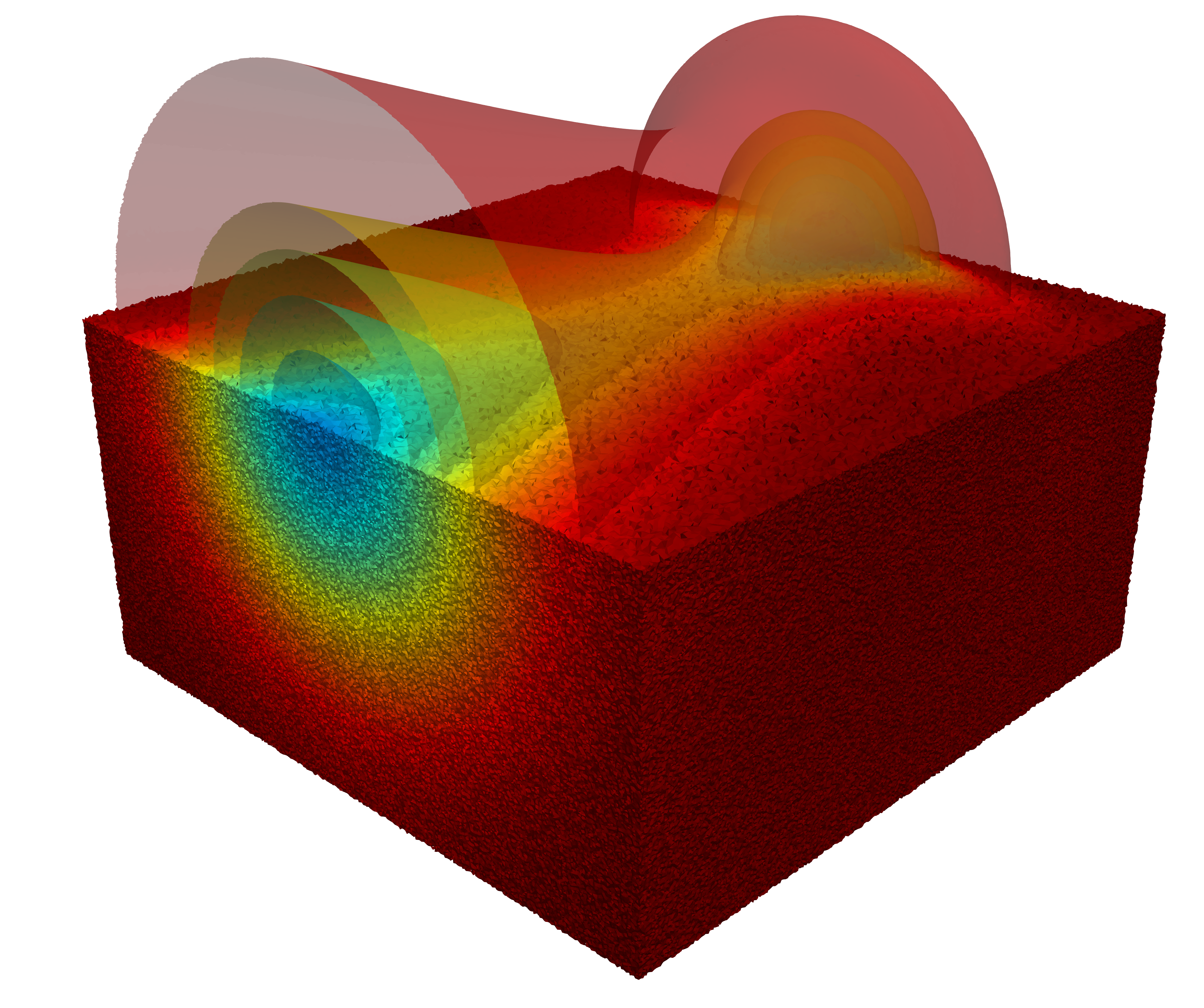}
		\caption{}
		\label{fig:podm-mesh-contours}
	\end{subfigure}
	\caption{Stability data and visualization of the generated mesh of this use-case.
	(a): Comparison of the minimum dihedral angle between  the different back-ends.
	(b): Visualization of the generated mesh along with contours of the dataset.}
	\label{fig:podm-stability}
\end{figure}

Similar to Section~\ref{sec:cdt3d-stability}, 
Figure~\ref{fig:podm-quality} compares a mesh quality measure 
(minimum dihedral angle in this case) among the different tasking approaches
of this case-study. In order to satisfy the \emph{stability} 
requirement, the quality among the different execution back-ends
should be comparable.
The histograms of Figure~\ref{fig:podm-quality} are built using the 
meshes generated at $40$ cores in the previous section and averaging 
the data over the $10$ runs of the experiment.
The difference between the tasking approach and the baseline is
marginal. The deviation from the baseline is lower that of
the previous case-study due to the different meshing method used.

\section{Conclusion}
In this work, we presented a high level tasking framework and applied it to a 
number of different meshing operations employing a speculative approach. 
A generic solution like the one presented in this 
work would be expected to introduce some overhead 
compared to hand-optimized code written for the 
specific needs of the application. However, such an 
overhead is only made visible in the second case study
(PODM, Section \ref{sec:case-study-PODM}), 
which is affected to a degree by the tighter 
integration of thread management with the application 
algorithm. 
In contrast, the first benchmark
(CDT3D, Section  \ref{sec:case-study-CDT3D}) exhibits 
performance improvements thanks to the well-optimized scheduling 
algorithm of the tasking framework.
These improvements are significantly higher when compared 
to the straight-forward use of tasks 
and higher than 
higher-level constructs that are already present in some of
the back-ends 
(\ompfor, \omptaskloop, \tbbfor).
\autoref{tab:total-time} indicates that 
the use of the provided higher-level constructs or the \texttt{flat} strategy which corresponds to the 
straight-forward use of tasks, introduces overheads when compared to the baseline application.
The naive \texttt{flat} strategy performs the worst between the two since it  produces too many tasks and in a sequential fashion.
\texttt{omp-flat} for example, results in an almost $1200\%$ slowdown. 
Introducing the \texttt{hierarchical} strategy that creates tasks recursively
reduces the overhead of creating tasks by distributing it among different threads.
This approach offers small gains in low number of cores and up to $4.39\%$ 
improvement for the total running time when
utilizing the Argobots back-end. The best results were obtained with 
the \texttt{2level} strategy, a hybrid of the other two, that manages
improvements of up to $2.31\%$ at $10$ cores and $5.81\%$ at $40$ cores
when compared to the baseline application.  
When it comes to individual operations, Tables \ref{tab:point-creation-local-reconnection-table} and 
\ref{tab:edge-collapse-smoothing} indicate performance improvements of up to
$2.04\%$ for the operations that use application-specific scheduler in the baseline implementation 
and up to $13\%$ for the mesh operations that use generic OpenMP constructs in the baseline application.

Moreover, the abstract front-end gives a platform to explore multiple execution back-ends; 
Figures \ref{fig:high-level-vs-flat} and \ref{fig:flat-vs-two-level-vs-hierarchical}
show results over $12$ different \texttt{strategy-backend} 
combinations, which are accessible to 
the application developer through a compile-time parameter.
In terms of stability, Sections \ref{sec:cdt3d-stability} and \ref{sec:podm-stability} indicate that the
introduction of the tasking framework has no effect on the quality of the generated data.

Finally, separating the concerns of \emph{functionality} and 
\emph{performance} is a crucial step towards the implementation of the
\emph{Telescopic Approach}~\cite{chrisochoides_telescopic_2016},
which lays down a design
for achieving scalability for mesh generation on exascale machines. 
The  generalized tasking framework facilitates the integration with the PREMA 
runtime system~\cite{Thomadakis20M}  at the shared memory level by handing 
control of thread management and load balancing from the application to the 
runtime system. This decoupling is expected to speedup the 
implementation due to the improved encapsulation of the different methods and PREMA's more efficient management of hardware resources. 
For example, the application independent tasking pools that the tasking framework offers
can provide load balancing across different instances of the same application occupying 
a common shared memory space. This scenario fits well with our previous
work~\cite{feng_scalable_2017} and the 
Parallel Data Refinement layer of the \emph{Telescopic Approach}.

The separation of concerns with regard to functionality and performance
allows to create reusable modules for future applications.
By extracting good solutions from previous implementations,
we can reduce the implementation effort of future revisions of our applications but also
provide templates for developing new speculative methods.

\section{Future Work}

Both case studies introduce a parameter (\emph{grainsize} for CDT3D and 
\emph{schedule\_limit} for PODM) that controls the number of tasks created.
In this study, we determined the optimal values by scanning over a predefined range.
However, a more thorough study that includes different meshing inputs and parameters
is needed in order to identify a set of optimal values that exhibits the best
performance on average across different inputs.
As with any parameter optimization study, this could be performed 
utilizing machine learning on pre-generated data. Moreover, the  underlying 
tasking framework could be equipped with an 
online machine learning method~\cite{hoi_online_2018} that can choose the optimal parameters based on runtime data.
Based on previous experience \cite{barker_practical_2005}, we believe that an online
machine learning method is superior to formal load balancing descriptions which tend to
oversimplify the problem when it comes to the effect of 
different hardware and problem configurations to the
application. On the other hand, an online machine learning framework could be
trained on multiple datasets of interest and  reinforce its models based on
runtime data.

Utilizing tasks in conjunction with the speculative approach could be further 
improved by abstracting the Parallel Correctness sections of
Figure~\ref{fig:codes}. This could be achieved with high level (but still 
application-specific) abstractions that lock and unlock the cavity of an 
operation 
automatically. 
The addition of contention managers and the ability to automatically reschedule
aborted speculative tasks will make this framework more complete and improve
the encapsulation by providing to the user a tasking framework that is able to
decouple all three sections of Figure~\ref{fig:codes}. Moreover, the decoupling of
the three responsibilities  will allow to refactor both our case-studies 
and extract a larger number of abstract parameters that control scheduling decisions.
Studying the effect of these additional parameters is part of our future work.

The back-end of the \texttt{task\_for} front-end of Listing~\ref{code:task_for} is 
currently a compile-time parameter. However, there is no technical constrain that
would prevent it from being an extra argument of the front-end API. 
Making the back-end a parameter will enable interoperability among the different
back-ends based on the needs of the user.

Taking into account the memory affinity layout when scheduling 
tasks could also improve the framework's performance. As part of our future work
for the Argobots back-end, we intend to examine this aspect in more detail.
Specifically, the back-end will utilize the Linux \texttt{numa} library to collect memory
related information about the underlying hardware. This information will then be
used by the back-end to coordinate task stealing attempts between threads,
in a hierarchical way. 

Throughout this study the default load balancing/work scheduler of each back-end was used.
This work could be extended by evaluating the available work schedulers of OpenMP
(\texttt{static, dynamic, guided}) and implementing \emph{breadth-first} and \emph{work-first}
strategies \cite{hutchison_evaluation_2008}  for the task-based methods. Moreover, 
we could customize the \texttt{tbb} scheduler and extend our Argobots back-end with custom load
balancing methods. Having a variety of scheduling methods for each back-end  will allow
to produce a similar study based on the effect of the application-specific
scheduling algorithms of our two case-studies versus generic load balancing methods.

In a follow-up study, we plan to explore more back-end systems that can utilize 
both homogeneous and heterogeneous platforms, including GPUs. Generic
heterogeneous frameworks such as SYCL 
\footnote{\url{https://www.khronos.org/sycl} (Accessed 27th April 2021)}
and Kokkos
\footnote{\url{https://kokkos.org} (Accessed 27th April 2021)}
provide already support for launching and managing tasks on GPUs.
Combining them  with the tasking framework is expected to assist
in hiding latencies related to data transfers to and from the
device, as well as delays launching kernels.
Evaluating such a framework would also require the addition or extension
of current mesh operations for heterogeneous architectures.

\begin{acknowledgements}
We would like to thank the reviewers 
for providing helpful comments on earlier drafts of the manuscript.
This research was sponsored in part by
the NASA Transformational Tools and
Technologies Project (NNX15AU39A) of the 
Transformative Aeronautics Concepts Program under
the Aeronautics Research Mission Directorate, 
NSF grant no. CCF-1439079, the Richard T.
Cheng Endowment, the Modeling and Simulation fellowship of 
Old Dominion University and the Dominion Scholar fellowship of Old 
Dominion University.
Experiments were supported by the Research Computing clusters at Old Dominion 
University. The authors would like to thank Kevin Garner 
for the corrections of the English text in the manuscript.
\end{acknowledgements}

\bibliography{mybibfile}

\end{document}